\documentclass[preprint, times]{aastex62}
\usepackage{float}
\usepackage{ulem}

\newcommand\msunyr{\rm {\it M}_{\odot}\,yr^{-1}}
\newcommand\mum{$\mu$m}
\newcommand\mums{$\mu$m }

\begin{document}

\title{Disk Masses and Dust Evolution of Protoplanetary Disks Around Brown Dwarfs}
\author[0000-0002-3091-8061]{Anneliese M. Rilinger}
\affiliation{Department of Astronomy and Institute for Astrophysical Research, Boston University, 725 Commonwealth Avenue, Boston, MA 02215}

\author[0000-0001-9227-5949]{Catherine C. Espaillat}
\affiliation{Department of Astronomy and Institute for Astrophysical Research, Boston University, 725 Commonwealth Avenue, Boston, MA 02215}

\accepted{June 28, 2021 to ApJ}
\correspondingauthor{Anneliese M. Rilinger}
\email{amr5@bu.edu}

\begin{abstract}
We present the largest sample of brown dwarf (BD) protoplanetary disk spectral energy distributions modeled to date.  We compile 49 objects with ALMA observations from four star-forming regions: $\rho$ Ophiuchus, Taurus, Lupus, and Upper Scorpius.  Studying multiple regions with various ages enables us to probe disk evolution over time.  Specifically, from our models we obtain values for dust grain sizes, dust settling, and disk mass; we compare how each of these parameters vary between the regions.  We find that disk mass decreases with age.  We also find evidence of disk evolution (i.e., grain growth and significant dust settling) in all four regions, indicating that planet formation and disk evolution may begin to occur at earlier stages.  We generally find these disks contain too little mass to form planetary companions, though we cannot rule out that planet formation may have already occurred.  Finally, we examine the disk mass -- host mass relationship and find that BD disks are largely consistent with previously-determined relationships for disks around T Tauri stars.
\end{abstract}

\keywords{brown dwarfs - planets and satellites: formation - protoplanetary disks - stars: formation}

\section{Introduction}
Young, newly-formed stars are commonly surrounded by disks of gas and dust, from which planetary companions may form \citep{williams11}. Brown dwarfs (BDs) have also been observed to host these protoplanetary disks \citep{comeron98, natta01, muench01, natta02}.  Though usually smaller in size \citep{hendler17}, BD protoplanetary disks are similar to the disks around their young stellar counterparts (e.g., T Tauri stars, TTS) in many ways.  Both BD disks and TTS disks are typically flat \citep{scholz07} with evidence of dust grain growth in their inner regions \citep{sterzik04, apai05, meru13, pinilla13}.  Some BD disks also show signs of inner disk clearing \citep{muzerolle06, rilinger19}.  Furthermore, the fraction of BDs observed to host disks is similar to the TTS disk fraction, possibly indicating a shared formation mechanism \citep{luhman05}.

Far-infrared surveys of BDs with the \textit{Herschel} space observatory \citep[e.g.,][]{harvey12, liu15, daemgen16} greatly enhanced our understanding of BD disks.  Including \textit{Herschel} far-IR and sub-millimeter photometry in BD SEDs enabled better constraints on disk parameters via radiative transfer modeling.  Such models show similar disk structure and geometry to T Tauri disks, including disk scale height \citep{liu15}, and dust settling, though BD disks may be even more settled than their TTS counterparts \citep{daemgen16}.  BD disk masses calculated from far-IR and sub-mm fluxes are consistently lower than typical TTS disk masses, which is consistent with the positive correlation between disk mass and host mass reported by, e.g., \citet{andrews13} and \citet{pascucci16}.  However, as noted by \citet{daemgen16}, the far-IR flux mostly traces the shape and temperature of the disk, while (sub)-mm fluxes are more sensitive to the disk mass.

Fortunately, comprehensive millimeter-wavelength surveys of protoplanetary disks have been made possible with ALMA.  Such surveys have been performed in various star-forming regions for both TTS hosts \citep[e.g.,][]{andrews13, ansdell16, barenfeld16, eisner16, pascucci16, ansdell17, eisner18, ruiz-rodriguez18, cazzoletti19, cieza19, vanterwisga19, vanterwisga20, grant21} as well as BD hosts in Ophiuchus, \citep{testi16}, Upper Scorpius \citep{vanderplas16}, Taurus \citep{ward-duong18}, and Lupus \citep{sanchis20}. Though comparisons have been made between the TTS protoplanetary disks in the different regions, no study has yet been performed to assess how the BD disks vary across regions.  

Disk mass, the amount of gas and dust in the disk, is one of the most important disk parameters to measure.  Protoplanetary disks serve as the mass reservoir for planet formation and are thought to be the sites where planet formation occurs \citep{apai05}; thus, constraining the amount of material in the disk directly informs the potential for planet formation. BDs have been observed to host planetary-mass candidates \citep{han13, udalski15, shvartzvald17}.  Since planets are known to be common around M dwarf stars \citep{dressing13, bonfils13}, it follows that planets may also be common around BDs.  The formation mechanism of the BDs themselves, however, may affect their ability to form planetary companions.  BD formation mechanisms can be categorized into two groups: non-ejection and ejection.  In the non-ejection scenario, no significant disruptions affect the condensation of the BD from its mass reservoir \citep{hennebelle09, andre12, riaz18}.  On the other hand, an impulsive reaction with one or more nearby objects may abruptly eject the BD out of its mass reservoir \citep{reipurth01, bate03, basu12, stamatellos09}.  In these latter ejection scenarios, the disk around the BD is expected to be truncated or even stripped completely, reducing the amount of material available for planet formation.  Simulations indicate that a BD disk requires approximately one Jupiter mass of material in order to form a planet on the order of a few Earth masses; the mass of potential planets and their likelihood of forming decrease as the mass of the disk decreases \citep{payne07}.

As disks evolve, dust grains are expected to grow and settle: the first steps towards forming planetesimals and eventually planets.  Small ($\sim$0.25 \mum) amorphous silicate dust grains originating in the ISM collide with one another and can grow to $\sim$micron sizes if the collisions are inelastic.  As grains grow larger, they become more susceptible to the drag force they experience as they move through the gas in the disk; the dust grains thus experience the pull of gravity toward the disk midplane and settle out of the disk atmosphere \citep{dullemond04}.  Observational studies have shown that grain growth and dust settling are already underway by $\sim$1 Myr \citep{furlan09, ribas17, grant18}, though older disks typically show more significant evidence of this evolution.

In addition to the vertical motion of dust grains settling to the midplane, dust grains also drift radially inwards due to the sub-Keplerian motion of the gas and the subsequent ``headwind'' experienced by the dust \citep{whipple72, weidenschilling77}.  This radial drift is predicted to be more pronounced for BD disks than TTS since the gas velocity deviates more from the Keplerian speed as stellar mass decreases \citep{pinilla13, pinilla17}.  However, mm-wavelength observations and measurements of the millimeter spectral index of BD disks show evidence of large dust grains present in these disks \citep[e.g.,][]{ricci13, ricci14}, indicating that some other process prevents radial drift from depleting the disks of large grains.  Possible solutions to the radial drift problem include pressure bumps within the disk, potentially created by planetary companions \citep{pinilla12, pinilla14, espaillat14, owen16}, lower gas masses than expected within the disks \citep{pinilla17}, or growth of fluffy aggregates, which would remain coupled to the gas \citep{kataoka13}.  Evidence of structure (e.g., rings and gaps) within BD disks could lend support to the pressure bump theory.  On the other hand, fluffy aggregates may have similar opacities to compact dust grains at mm wavelengths and may be difficult to distinguish from mm-sized dust \citep{kataoka14}.

In this work, we compile the largest sample of BD disks modeled to date, drawn from the aforementioned studies, in order to search for similarities and differences between BD protoplanetary disks in various star-forming regions.  We focus on four regions: Ophiuchus \citep[0.5-2 Myr, $\sim$140 pc,][]{wilking08, ortizleon17}, Taurus \citep[1-2 Myr, $\sim$140 pc,][]{kh95, luhman10, andrews13}, Lupus \citep[1-3 Myr, 150-200 pc,][]{comeron08, ansdell16}, and Upper Scorpius \citep[5-11 Myr, $\sim$145 pc,][]{preibisch02, pecaut12}.  In particular, we compare disk masses and disk evolution (characterized by grain growth and dust settling) in these regions.  By considering regions of varying age, we can probe how BD disks change over their lifetimes.

In Section \ref{sample} we present our BD sample selection.  Our modeling analysis is described in Section \ref{analysis}.  We discuss the results of our analysis in Section \ref{discussion}, and summarize our findings in Section \ref{conclusion}.

\section{Sample}\label{sample}
We have compiled a sample of 49 BDs with protoplanetary disks located in the $\rho$ Ophiuchus, Taurus, Lupus, and Upper Scorpius star-forming regions.  This constitutes the largest sample of BD disk spectral energy distributions (SEDs) modeled to date.  These four regions were selected because the BDs in these regions have been observed in the (sub-)millimeter: Ophiuchus \citep[][and Testi et al., submitted]{testi16}, Taurus \citep{andrews13, ward-duong18}, Lupus \citep{sanchis20}, and Upper Sco \citep{vanderplas16}.  We only consider objects with masses less than the hydrogen-burning limit ($\sim$0.08 M$_{\odot}$) and/or spectral types of M6 or later since these objects are most likely to be bona fide BDs \citep{luhman05}.  We also required the objects to have millimeter-wavelength photometry measurements or upper limits, in order to be able to better constrain our SED models.  Additionally, we did not include objects in our sample which are known to be in binary or multiple systems, since companions can affect disk properties. Finally, one object, 2MASS J04381486+2611399, has been reported to be a highly-inclined, edge-on disk \citep{luhman07} so we remove this object from our sample.  Table \ref{table:sample} lists the 49 objects in our final sample, along with their spectral types, temperatures, extinctions, luminosities, masses, radii, and distances, compiled from the literature (see references in Table \ref{table:sample}).  We note that the relevant properties have been scaled to Gaia DR2 distances for all objects that had Gaia DR2 parallaxes \citep{gaia16, gaia18}.

\startlongtable
\begin{deluxetable*}{c c c c c c c c c}
\tablecaption{Parameters of BDs modeled in this work\label{table:sample}}
\tablehead{
\colhead{Object} & \colhead{SpT}
 & \colhead{T$_*$} & \colhead{A$_{V}$} & \colhead{L$_*$} & \colhead{M$_*$} & \colhead{R$_*$} & \colhead{Distance} & \colhead{Source}\\  & & \colhead{(K)} & & \colhead{log(L$_{\odot}$)} & \colhead{(M$_{\odot}$)} & \colhead{(R$_{\odot}$)} & \colhead{(pc)} & 
}
\startdata
\multicolumn{9}{c}{Members of Taurus}\\
\tableline
J04390396+2544264	& M7.25 & 2837 & 0.5 & -1.336 & 0.035 & 0.552 & 143.99$^{+4.50}_{-4.24}$ & 1\\
J04141188+2811535 & M6.25 & 2963 & 2.5 & -1.746 & 0.053 & 0.873 & 131.09$^{+2.92}_{-2.79}$ & 1\\
J04292165+2701259 & M6 & 3091 & 2 & -0.115 & 0.058 & 1.884 & 140\tablenotemark{a} & 1\\
J04230607+2801194 & M6 & 2990 & 1.5 & -1.332 & 0.058 & 0.942 & 133.88$^{+2.49}_{-2.40}$ & 1\\
KPNO Tau 3 & M6 & 2990 & 1.6 & -1.655 & 0.058 & 0.377 & 155.88$^{+5.72}_{-5.33}$ & 1\\
GM Tau & M6.5 & 2935 & 0.6 & -1.507 & 0.048 & 0.794 & 138.31$^{+2.91}_{-2.79}$ & 1\\
J04390163+2336029 & M6 & 2990 & 0.5 & -1.054 & 0.058 & 1.13 & 127.81$^{+1.31}_{-1.29}$ & 1\\
J04400067+2358211 & M6 & 2858 & 0.5 & -1.566 & 0.058 & 0.67 & 120.49$^{+2.35}_{-2.26}$ & 1\\
J04414825+2534304 & M7.75 & 2752 & 1.3 & -1.683 & 0.028 & 0.634 & 136.16$^{+3.84}_{-3.64}$ & 1\\
J04442713+2512164 & M7.25 & 2838 & 0.0 & -1.553 & 0.05 & 0.69 & 141.0$\pm$2.7 & 2\\
CFHT Tau 4 & M7 & 2880 & 5.67 & -0.757 & 0.095 & 1.68 & 147.1$\pm$5.1 & 2\\
CFHT Tau 9 & M5.75 & 3023 & 2.14 & -1.517 & 0.064 & 0.636 & 155.51$^{+3.21}_{-3.08}$ & 3\\
KPNO Tau 6 & M8.5 & 2555 & 1.28 & -2.468 & 0.02 & 0.298 & 116.05$^{+7.65}_{-6.76}$ & 3\\
KPNO Tau 7 & M8.25 & 2632 & 1.57 & -2.272 & 0.024 & 0.352 & 122.82$^{+7.35}_{-6.57}$ & 3\\
J04330945+2246487 & M6 & 2990 & 4.53 & -1.31 & 0.069 & 0.825 & 148.98$^{+10.06}_{-8.87}$ & 3\\
J04214631+2659296 & M5.75 & 3023 & 4.72 & -1.529 & 0.063 & 0.627 & 160.46$^{+7.63}_{-6.97}$ & 3\\
J04414489+2301513 & M8.5 & 2555 & 1.4 & -2.262 & 0.023 & 0.377 & 120.42$^{+5.74}_{-5.24}$ & 3\\
KPNO Tau 12 & M9 & 2400 & 1.88 & -2.973 & 0.011 & 0.189 & 140\tablenotemark{a} & 3\\
J04201611+2821325 & M6.5 & 2935 & 2.3 & -1.87 & 0.041 & 0.449 & 128.99$^{+3.43}_{-3.25}$ & 3\\
J04263055+2443558 & M8.75 & 2478 & 0.9 & -2.26 & 0.02 & 0.402 & 119.88$\pm10.05$ & 3\\
J04290068+2755033 & M8.25 & 2700 & 1.71 & -1.98 & 0.023 & 0.467 & 145.65$^{+7.85}_{-7.08}$ & 3\\
J04554969+3019400 & M6 & 2990 & 0.09 & -1.695 & 0.052 & 0.529 & 156.51$^{+4.76}_{-4.49}$ & 3\\
J04574903+3015195 & M9.25 & 2350 & 0.2 & -2.96 & 0.015 & 0.2 & 140\tablenotemark{a} & 4\\
\tableline
\multicolumn{9}{c}{Members of Ophiuchus}\\
\tableline
CFHTWIR-Oph 16 & M8 & 2710 & 11.71 & -2.00 & 0.03 & 0.45 & 139.4\tablenotemark{a} & 6\\
ISO-Oph 23 & M6.5 & 2930 & 7.91 & -1.40 & 0.07 & 0.69 & 139.4\tablenotemark{a} & 5\\
ISO-Oph 32 & M7.25 & 2840 & 2.09 & -1.18 & 0.08 & 1.06 & 151.26$^{+4.88}_{-4.58}$ & 5\\
ISO-Oph 33 & M87 & 2880 & 6.82 & -2.23 & 0.06 & 0.31 & 139.4\tablenotemark{a} & 5\\
GY92 141 & M8.5 & 2550 & 1.16 & -2.54 & 0.02 & 0.28 & 142.7\tablenotemark{a} & 6\\
CFHTWIR-Oph 58 & M8 & 2710 & 6.20 &	-2.7 & 0.04 & 0.2 &		139.4\tablenotemark{a} & 6\\
CFHTWIR-Oph 66 & M7.75 & 2750 &	10.62 & -2.65 & 0.05 & 0.21 & 139.4\tablenotemark{a} & 6\\
CFHTWIR-Oph 77 & M9.75 & 2260 &	7.29 & -3.02 & 0.02	& 0.2	& 139.4\tablenotemark{a} & 6\\
GY92 264 & M8 & 2710 & 0.31 & -1.74 & 0.04 & 0.61 & 139.15$^{+2.88}_{-2.77}$ & 5\\
CFHTWIR-Oph 90 & L0 & 2250 & 1.94 & -2.65 & 0.01 & 0.31	& 139.4\tablenotemark{a} & 6\\
GY92 320 & M7.5 & 2753 & 2.09 & -2.15 & 0.04 & 0.37 & 140.03$^{+15.11}_{-12.43}$ & 5\\
CFHTWIR-Oph 98 & M9.75 & 2260 & 3.72 & -2.96 & 0.01 & 0.22 & 139.4\tablenotemark{a} & 6\\
CFHTWIR-Oph 107 & M6.25 & 2960 & 1.71 & -2.02 & 0.07 & 0.37 & 159.6\tablenotemark{a} & 6\\
\tableline
\multicolumn{9}{c}{Members of Lupus}\\
\tableline
J15451851-3421246 & M6.5 & 2935 & 0 & -1.40 & 0.09 & 0.774 & 152$\pm$4 & 7\\
SONYC Lup 3-7 &	M8.5 & 2600	& 0 & -2.0 & 0.02 & 0.493 & 150$\pm$6 & 7\\
Lup 706	& M7.5 & 2795 & 0 & -2.70 & 0.05 & 0.191 & 158.5\tablenotemark{a} & 7\\
AKC 2006-18	& M6.5 & 2935 & 0 & -2.0 & 0.07	& 0.387	& 149$\pm$8 & 7\\
Lup 818s & M6 & 2990 & 0 & -1.70 & 0.09	& 0.527 & 157$\pm$3 & 7\\
J16101984-3836065 & M6.5 & 2935 & 0 & -1.40 & 0.09 & 0.774 & 159$\pm$3 & 7\\
J16085529-3848481 & M6.5 & 2935 & 0.5 & -1.30 & 0.09 & 0.865 & 158$\pm$3 & 7\\
Lup 607 & M6.5 &	2935 & 0 & -1.30 & 0.1 & 0.865 & 175$\pm$6 & 7\\
\tableline
\multicolumn{9}{c}{Members of Upper Scorpius}\\
\tableline
J15555600-2045187 & M6.5 & 3400 & 3.38 & -1.56 & 0.07 & 0.479 & 145.65$^{+4.66}_{-4.38}$ & 8\\	
J15560104-2338081 & M6.5 & 2800 & 1.3 & -1.99 & 0.07 & 0.432 & 139.88$^{+5.65}_{-5.23}$ & 8\\	
J15591135-2338002 & M7 & 2500 & 1.3 & -2.20 & 0.06 & 0.421 & 139.94$^{+7.39}_{-6.69}$ & 8\\	
J16100541-1919362 & M7 & 2600 &	0.7 & -2.19 & 0.06 & 0.397 & 149.69$^{+7.40}_{-6.73}$ & 8\\
J16060391-2056443 & M7.5 & 2800 & 1.5 & -1.83 &	0.04 & 0.518 & 137.15$^{+5.22}_{-4.85}$ & 8\\
\enddata
\tablenotetext{a}{From reference listed in Source column.}
\tablecomments{Distances were computed from Gaia parallaxes \citep{gaia16, gaia18} unless otherwise noted. These distances were used to scale the relevant properties obtained from the literature.} The Source column gives the reference for all other properties listed here: (1) \citet{ward-duong18}; (2) \citet{ricci14} and references therein; (3) \citet{andrews13}; (4) \citet{luhman04}; (5) \citet{testi16} and Testi et al., submitted; (6) Testi et al., submitted; (7) \citet{sanchis20}; (8) mass and spectral type from \citet{vanderplas16}, other properties from \citet{liu15}.
\end{deluxetable*}

\section{Analysis and Results}\label{analysis}
\subsection{SED Modeling}\label{sedmodeling}
The disk properties for each object were determined by fitting disk structure models to the SED of each BD.  To construct the SEDs, photometric data points from visible to millimeter wavelengths were taken from the Vizier catalogue access tool \citep{vizier}.  Of the 49 objects in this sample, 31 also have low resolution (R $\sim$ 60--130) spectra from the InfraRed Spectrograph (IRS) on the Spitzer Space Telescope \citep{houck04}.  The SEDs for each object are presented in Appendix \ref{appendix:models}, Figures \ref{fig:all_tau_seds}, \ref{fig:all_oph_seds}, \ref{fig:all_lup_seds}, and \ref{fig:all_sco_seds}.

We use the D'Alessio Irradiated Accretion Disk (DIAD) radiative transfer models \citep{diad98, diad99, diad01, diad05, diad06} to model the vertical and radial structure of the disks.  Previous works \citep[e.g.,][]{morrow08, adame11, rilinger19} have used these models to fit BD SEDs.  The disks are assumed to be irradiated by the central BD.  DIAD calculates the vertical and radial structure of each disk self-consistently while enforcing hydrostatic equilibrium.  Two populations of dust grains are modeled: a population of small dust grains in the atmosphere of the disk and another consisting of large grains in the disk midplane.  In each dust population, the minimum grain size is fixed at 0.005 \mum; particle sizes are distributed following a power law with a power of --3.5 \citep{mathis77} up to maximum sizes that are left as free parameters in the model (a$_{max, atm}$ and a$_{max, mid}$).  The disk mass is calculated by integrating over the surface mass density ($\Sigma$), which in turn is determined by the mass accretion rate ($\dot{M}$) and disk viscosity ($\alpha$; $\Sigma \propto \dot{M} \alpha^{-1}$). An important feature of the DIAD models is that $\Sigma$ is calculated self-consistently with the vertical and radial structure of the disk, as opposed to assuming a fixed power-law surface density distribution. Of the 49 objects in this sample, 18 have measured $\dot{M}$ values in the literature (see Table \ref{table:mdot}).  We fix $\dot{M}$ for these 18 objects; for the remaining objects we adopt a value of $1\times10^{-10}$ ${\msunyr}$, which is approximately the average of the 18 measured $\dot{M}$ values. Disk viscosity ($\alpha$) is left as a free parameter in the DIAD model.  

\begin{deluxetable}{c c c}
\tablecaption{BDs with literature accretion rates\label{table:mdot}}
\tablehead{
\colhead{Object} & \colhead{$\dot{M}$} & \colhead{Source}\\  & (M$_{\odot}$ yr$^{-1}$) & 
}
\startdata
J04390396+2544264 & 6.9 x 10$^{-12}$ & 1\\
J04141188+2811535 & 9.0 x 10$^{-10}$ & 1\\
GM Tau & 1.9 x 10$^{-9}$ & 1\\
J04414825+2534304 & 1.9 x 10$^{-11}$ & 1\\
J04442713+2512164 & 2.0 x 10$^{-10}$ & 2\\
KPNO Tau 6 & 3.6 x 10$^{-11}$ & 1\\
KPNO Tau 12 & 1.6 x 10$^{-11}$ & 1\\
ISO-Oph 23 & 1.4 x 10$^{-10}$ & 3\\
ISO-Oph 32 & 3.2 x 10$^{-11}$ & 3\\
ISO-Oph 33 & 2.0 x 10$^{-11}$ & 3\\
J15451851-3421246 & 1.77 x 10$^{-11}$ & 4\\
SONYC Lup 3-7 & 1.39 x 10$^{-11}$ & 4\\
Lup 706 & 1.52 x 10$^{-12}$ & 4\\
AKC 2006-18 & 5.48 x 10$^{-12}$ & 4\\
Lup 818s & 1.15 x 10$^{-11}$ & 4\\
J16101984-3836065 & 2.74 x 10$^{-11}$ & 4\\
J16085529-3848481 & 1.91 x 10$^{-11}$ & 4\\
Lup 607 & 3.34 x 10$^{-12}$ & 4\\
\enddata
\tablecomments{Sources: (1) \citet{herczeg08}; (2) \citet{bouy08}; (3) \citet{manara15}; (4) \citet{alcala17}, scaled to Gaia distances.}
\end{deluxetable}

The inner edge or ``wall'' of the disk is set by T$_{wall,in}$, a temperature that is a free parameter in the DIAD model.  For a full-disk model, T$_{wall,in}$ is set to the dust destruction temperature, which we assume to be 1400 K.  To model a transitional disk (i.e., a disk with a hole), this temperature can be decreased, effectively moving the inner edge of the disk further from the central BD.  We can also model a pre-transitional disk (i.e., a disk with a large gap separating the inner and outer disk) by modifying the disk structure to include an annular gap within the disk \citep{espaillat11}; in this case, the disk has both an inner wall and an outer wall, set by temperature T$_{wall,out}$.  Other free parameters of the DIAD model include the inner wall scale height (H$_{wall,in}$), disk outer radius (R$_{out}$), inclination (i), and dust settling ($\epsilon$, see Section \ref{settling}).  If the disk is a pre-transitional disk, the scale height of the outer wall, H$_{wall,out}$, is also a free parameter in the model.  We assume a fixed dust-to-gas mass ratio of 0.01.

\begin{figure*}
    \plotone{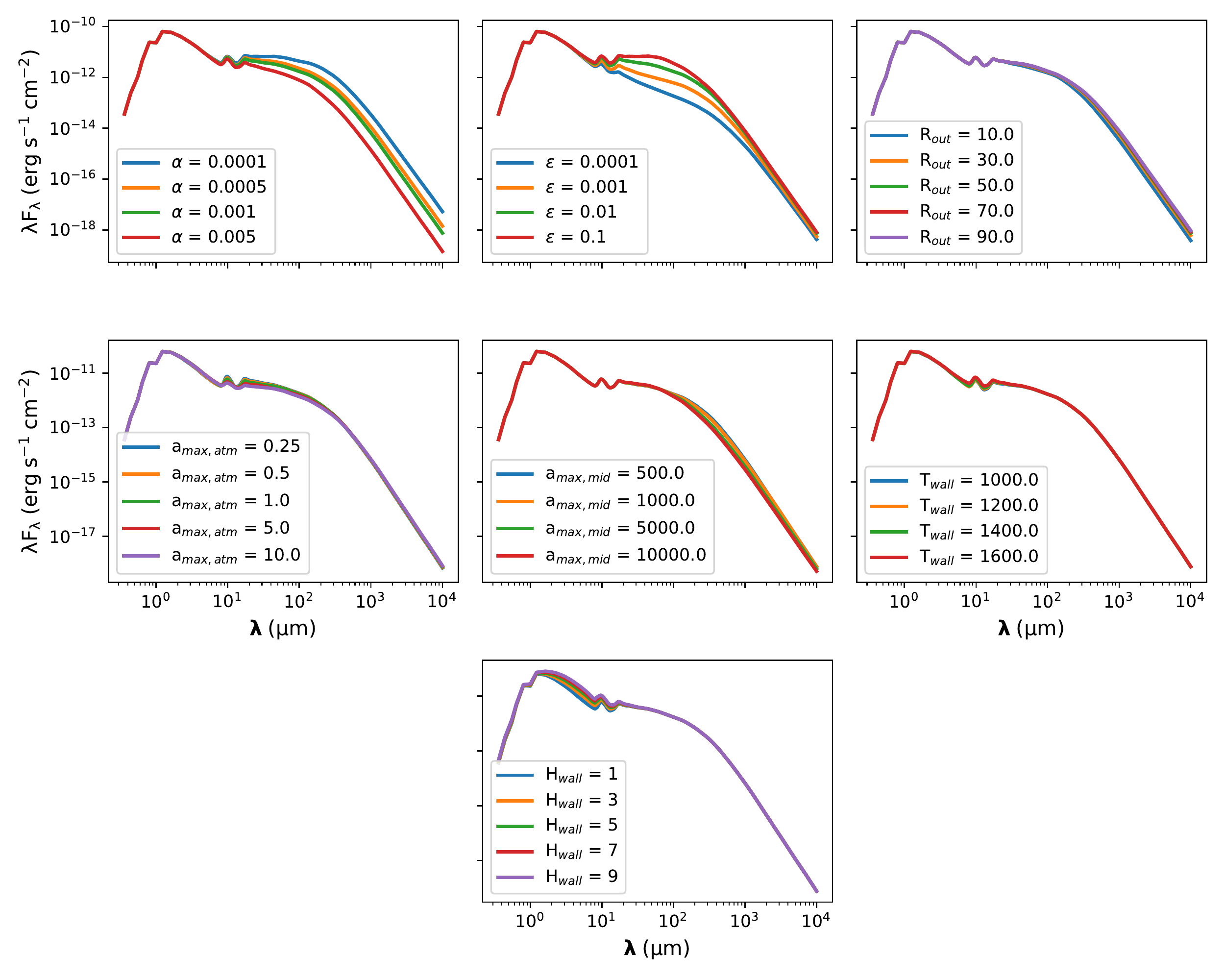}
    \caption{Effect of different parameters in the DIAD SED models.  In each panel, all but one parameter is held constant.  See text for the values of the fixed parameters.}
    \label{fig:paramtest}
\end{figure*}

In Figure \ref{fig:paramtest}, we explore the effect each of these disk parameters on the DIAD model.  All but one parameter is held constant in each panel to show how that parameter changes the shape of the resulting SED.  In all cases, we fix M$_*$, T$_*$, L$_*$, and distance to the average values from Table \ref{table:sample}.  We also fix inclination to 60$^{\circ}$, $\dot{M}$ to 1x10$^{-10}$ M$_{\odot}$ yr$^{-1}$, and the dust-to-gas mass ratio to 0.01.  The remaining default parameter values are as follows: $\alpha$ = 0.001, $\epsilon$ = 0.01, R$_{out}$ = 50 au, a$_{max, atm}$ = 1.0 \mum, a$_{max, mid}$ = 500 mm, T$_{wall}$ = 1400 K, and H$_{wall}$ = 1.  Figure \ref{fig:paramtest} shows the importance of including mm photometry in the SEDs: one or two mm-wavelength photometry points can help break the degeneracy between the $\alpha$ and $\epsilon$ parameters in the far-IR, resulting in more accurate measurements of the vertical structure (settling) as well as the disk mass.

Photospheres are constructed for each BD using the \citet{kh95} color table and the reported T$_*$, L$_*$, and R$_*$ values in Table \ref{table:sample}.  Colors are scaled to the dereddened observed J-band magnitude of the BD which is in turn calculated from the L$_*$ value in Table \ref{table:sample}.  We then interpolate to determine the photospheric emission at each wavelength.  Many objects in the sample have spectral types later than M6 (see Table \ref{table:sample}), the latest spectral type in the \citet{kh95} color table.  A spectral type of M6 was thus adopted for these objects.  Note that while the \citet{pecaut13} color table includes later spectral types, spectral types later than M5 are missing infrared wavelengths beyond K-band.  Since the two color tables show good agreement (within 10\% uncertainty) between the \citet{pecaut13} spectral type M7 colors and \citet{kh95} spectral type M6 colors, we opted to use the more complete \citet{kh95} color table to construct the BD photospheres.

DIAD disk models are computationally expensive to run: each model typically takes 1.5--2 hours to complete.  Therefore, it is unworkable to employ common statistical methods of estimating best-fit model parameters and their uncertainties (e.g., Markov-Chain Monte Carlo methods, Levenberg-Marquardt $\chi^2$ minimization, etc.).  To obtain best-fit SED models, we first ran a coarse grid of models for each object to explore the parameter space (see Table \ref{table:coarsegrid} for the initial grid parameters).  We then iteratively refined the grids as necessary based on visually inspecting the models with minimum $\chi^2$.  Upper limit points were not included in the $\chi^2$ calculation, but rather served as further visual guides for assessing the goodness of the fit (i.e., we do not accept a model that is inconsistent with the upper limits, even if it fits the other points well).  The reported values for the best-fit parameters thus represent qualitative approximations to the range of possible values of the parameters rather than accurate quantitative values.

\begin{deluxetable}{c c c}
\tablecaption{Initial Grid Parameters\label{table:coarsegrid}}
\tablehead{
\colhead{Parameter} & \colhead{Range} & \colhead{Units}
}
\startdata
$\epsilon$ & \{0.0001, 0.0005, 0.001, 0.005, 0.01, 0.05, 0.1\} & ...\\
$\alpha$ & \{0.0001, 0.0005, 0.001, 0.005, 0.01\} & ...\\
a$_{max, atm}$ & \{0.25, 1.0, 10.0\} & \mum\\
a$_{max, mid}$ & \{0.5, 1.0, 5.0, 10.0\}& mm\\
H$_{in}$ & \{1, 2, 3, 4, 5\} & ...\\
T$_{wall, in}$ & 400--1400, in steps of 100 & K\\
\enddata
\end{deluxetable}

In addition to the parameters in Table \ref{table:coarsegrid}, which were varied, three additional parameters were fixed for each object: $\dot{M}$, inclination, and R$_{out}$.  We fix $\dot{M}$ as described above.  Inclination was fixed at 60$^{o}$ except for J04442713 and CFHT Tau 4, whose inclinations were determined by \citet{rilinger19}.  Disk radii were adopted from the literature where possible.  We adopt the R$_{out}$ values determined via ALMA modeling by \citet{rilinger19} for J04442713 and CFHT Tau 4.  \citet{testi16} constrained the disk radii for the 5 BDs in their sample: the R$_{out}$ values for their BD disks are $\leq$ 25 au.  We adopt these R$_{out}$ values for those objects, and assume the same value for the additional 8 objects observed by Testi et al., submitted.  \citet{sanchis20} report disk radii for five of the disks in Lupus; we adopt these R$_{out}$ values rounded to the nearest ten.  For the remaining Lupus disks, we adopt the average of the $R_{out}$ values for this region, 50 au.  \citet{vanderplas16} did not resolve any of the disks in their Upper Sco sample.  An R$_{out}$ of 50 au is consistent with their angular resolution and the distance to these objects, and was adopted for consistency with the Lupus disks.  Similarly, \citet{ward-duong18} do not measure disk radii for the objects in their Taurus sample.  Adopting R$_{out}$ values of 50 au for these objects is consistent with our assumptions for other regions.  An R$_{out}$ of 50 au was assumed for the remaining disks in the sample, with two additional exceptions: the SEDs of J04330945 and J04263055 appear to show truncated disks.  R$_{out}$ values of 0.1, 0.2, 0.3, 0.4, 0.5, 1, 5, 10, and 15 au were tested for these two objects.  The best-fit SED models for each object are shown in Appendix \ref{appendix:models}, Figures \ref{fig:all_tau_seds}, \ref{fig:all_oph_seds}, \ref{fig:all_lup_seds}, and \ref{fig:all_sco_seds}.  The best-fit parameters for each of these models are given in Table \ref{table:params}.

\startlongtable
\begin{deluxetable*}{c c c c c c c c c c c}
\tablecaption{Best-fit parameters for BD SED models\label{table:params}}
\tablehead{
\colhead{Object} & \colhead{$\epsilon$} & \colhead{$\alpha$} & \colhead{a$_{max, atm}$} & \colhead{a$_{max, mid}$} & \colhead{z$_{wall, in}$} & \colhead{T$_{wall, in}$} & \colhead{R$_{wall, in}$\tablenotemark{a}} & \colhead{R$_{out}$}\tablenotemark{b} & \colhead{i}\tablenotemark{c} & \colhead{M$_{disk}$\tablenotemark{d}}\\  &  & & (\mum{}) & (mm) & (au) & (K) & (au) & (au) & ($^{\circ}$) & (M$_{Jup}$)}
\startdata
\multicolumn{11}{c}{Members of Taurus}\\
\tableline
J04390396+2544264 & 0.01 & 6 x 10$^{-5}$ & 10.0 & 0.5 & 0.0212 & 500 & 0.1 & 50 & 60 & 0.283\\
J04141188+2811535 & 0.0005 & 0.01 & 1.0 & 0.5 & 0.0531 & 400 & 0.25 & 50 & 60 & 0.188\\
J04292165+2701259 & 0.05 & 0.0005 & 0.25 & 0.5 & 0.0079 & 1400 & 0.06 & 50 & 60 &  0.366\\
J04230607+2801194 & 0.0005 & 0.0001 & 10.0 & 10.0 &  0.0011 & 1400 & 0.02 & 50 & 60 & 2.964\\
KPNO Tau 3 & 0.001 & 0.0001 & 10.0 & 10.0 & 0.0016 & 1400 & 0.01 & 50 & 60 & 2.964\\
GM Tau & 0.005 & 0.0005 & 10.0 & 5.0 & 0.0377 & 600 & 0.11 & 50 & 60 &  0.450\\
J04390163+2336029 & 0.0001 & 0.001 & 0.25 & 0.5 & 0.0016 & 1400 & 0.03 & 50 & 60 &  0.209\\
J04400067+2358211 & 0.001 & 0.0001 & 10.0 & 10.0 & 0.0005 & 1400 & 0.01 & 50 & 60 &  3.121\\
J04414825+2534304 & 0.01 & 7 x 10$^{-5}$ & 0.25 & 0.5 & 0.0007 & 1400 & 0.01 & 50 & 60 & 0.628\\
J04442713+2512164\tablenotemark{e}\tablenotemark{f} & 0.05 & 0.0005 & 1.0 & 5.0 &  0.0083 & 1400 & 0.07 & 100 & 40 & 2.050\\
CFHT Tau 4\tablenotemark{f} & 0.007 & 0.0007 & 0.25 & 3.0 & 0.0020 & 1400 & 0.07 & 80 & 70 & 0.420\\
CFHT Tau 9\tablenotemark{e} & 0.005 & 0.005 & 10.0 & 5.0 & 0.0006 & 1400 & 0.02 & 50 & 60 & 0.052\\
KPNO Tau 6 & 0.0005 & 0.0005 & 10.0 & 1.0 & 0.0004 & 1400 & 0.01 & 50 & 60 & 0.262\\
KPNO Tau 7 & 0.0005 & 0.0005 & 10.0 & 0.5 & 0.0008 & 1400 & 0.01 & 50 & 60 & 0.524\\
J04330945+2246487 & 0.0001 & 0.001 & 1.0 & 0.5 & 0.0010 & 1400 & 0.02 & 0.2 & 60 &  0.002\\
J04214631+2659296 & 0.01 & 0.005 & 10.0 & 10.0 & 0.0011 & 1400 & 0.02 & 50 & 60 &  0.052\\
J04414489+2301513 & 0.005 & 0.001 & 10.0 & 0.5 & 0.0003 & 1400 & 0.01 & 50 & 60 &  0.220\\
KPNO Tau 12 & 0.001 & 0.0005 & 1.0 & 0.5 & 0.0005 & 1400 & 0.003 & 50 & 60 &  0.095\\
J04201611+2821325 & 0.001 & 0.005 & 1.0 & 0.5 & 0.0005 & 1400 & 0.01 & 50 & 60 &  0.042\\
J04263055+2443558 & 0.1 & 0.005 & 1.0 & 1.0 & 0.0003 & 1400 & 0.01 & 0.1 & 60 &  1.37 x 10$^{-4}$\\
J04290068+2755033 & 0.001 & 0.001 & 1.0 & 1.0 & 0.0005 & 1400 & 0.01 & 50 & 60 &  0.157\\
J04554969+3019400 & 0.001 & 0.005 & 0.25 & 0.5 & 0.0011 & 1400 & 0.02 & 50 & 60 &  0.042\\
J04574903+3015195 & 0.1 & 0.0001 & 1.0 & 1.0 & 0.0001 & 1400 & 0.003 & 50 & 60 &  3.048\\
\tableline
\multicolumn{11}{c}{Members of Ophiuchus}\\
\tableline
CFHTWIR-Oph 16 & 0.05 & 0.001 & 1.0 & 5.0 & 0.0037 & 1400 & 0.01 & 25 & 60 & 0.157\\
ISO-Oph 23 & 0.001 & 0.001 & 10.0 & 10.0 & 0.0074 & 800 & 0.05 & 25 & 60 & 0.241\\
ISO-Oph 32 & 0.1 & 0.0005 & 10.0 & 1.0 & 0.0082 & 800 & 0.08 & 25 & 60 &  0.115\\
ISO-Oph 33 & 0.1 & 0.0002 & 0.25 & 10.0 & 0.0401 & 500 & 0.09 & 25 & 60 & 0.272\\
GY92 141 & 0.0005 & 0.001 & 10.0 & 5.0 & 0.0002 & 1400 & 0.004 & 25 & 60 & 0.131\\
CFHTWIR-Oph 58 & 0.01 & 0.005 & 10.0 & 5.0 & 0.0001 & 1400 & 0.003 & 25 & 60 & 0.036\\
CFHTWIR-Oph 66 & 0.1 & 0.0001 & 0.25 & 5.0 & 0.0011 & 1400 & 0.01 & 25 & 60 & 3.184\\
CFHTWIR-Oph 77 & 0.001 & 0.001 & 1.0 & 10.0 & 0.0001 & 1400 & 0.003 & 25 & 60 & 0.114\\
GY92 264 & 0.005 & 0.0001 & 1.0 & 10.0 & 0.0029 & 1400 & 0.01 & 25 & 60 & 1.56\\
CFHTWIR-Oph 90 & 0.05 & 0.005 & 10.0 & 1.0 & 0.0002 & 1400 & 0.004 & 25 & 60 & 0.018\\
GY92 320 & 0.005 & 0.005 & 10.0 & 10.0 & 0.0012 & 1400 & 0.01 & 25 & 60 & 0.031\\
CFHTWIR-Oph 98 & 0.1 & 0.001 & 10.0 & 10.0 & 0.0006 & 1400 & 0.003 & 25 & 60 & 0.135\\
CFHTWIR-Oph 107 & 0.0005 & 0.01 & 1.0 & 1.0 & 0.0003 & 1400 & 0.01 & 25 & 60 & 0.021\\
\tableline
\multicolumn{11}{c}{Members of Lupus}\\
\tableline
J15451851-3421246 & 0.01 & 5 x 10$^{-5}$ & 0.25 & 0.5 & 0.0164 & 1000 & 0.05 & 15 & 60 & 0.443\\
SONYC Lup 3-7 &	0.01 & 0.0001 & 10.0 & 10.0 & 0.0013 & 1400 & 0.01 & 30 & 60 & 0.199\\
Lup 706	& 0.05 & 5 x 10$^{-5}$ & 10.0 & 0.5 & 0.0045 & 700 & 0.02 & 95 & 60 & 0.268\\
AKC 2006-18	& 0.0005 & 0.0005 & 1.0 & 10.0 & 0.0003 & 1400 & 0.01 & 50 & 60 & 0.052\\
Lup 818s & 0.1 & 1 x 10$^{-5}$ & 1.0 & 5.0 & 0.0004 & 1400 & 0.01 & 20 & 60 & 2.138\\
J16101984-3836065 & 0.0005 & 0.0005 & 0.25 & 5.0 & 0.0007 & 1400 & 0.02 & 50 & 60 & 0.157\\
J16085529-3848481 & 0.0005 & 0.0001 & 10.0 & 5.0 & 0.0007 & 1400 & 0.02 & 40 & 60 & 0.578\\
Lup 607 & 0.0005 & 0.0001 & 1.0 & 5.0 & 0.0008 & 1400 & 0.02 & 50 & 60 &  0.147\\
\tableline
\multicolumn{11}{c}{Members of Upper Scorpius}\\
\tableline
J15555600-2045187 & 0.0005 & 0.001 & 10.0 & 10.0 & 0.0022 & 1400 & 0.02 & 50 & 60 & 0.325\\	
J15560104-2338081 & 0.0005 & 0.0005 & 10.0 & 1.0 & 0.0007 & 1400 & 0.01 & 50 & 60 & 0.733\\	
J15591135-2338002 & 0.001 & 0.0005 & 10.0 & 0.5 & 0.0009 & 1200 & 0.01 & 50 & 60 & 0.721\\	
J16100541-1919362 & 0.01 & 0.005 & 10.0 & 0.5 & 0.0012 & 1200 & 0.01 & 50 & 60 &  0.063\\
J16060391-2056443 & 0.0005 & 0.005 & 1.0 & 10.0 & 0.0005 & 1400 & 0.01 & 50 & 60 &  0.042\\
\enddata
\tablenotetext{a}{This is not a free parameter and was calculated using T$_{wall,in}$ following \citet{diad05}.}
\tablenotetext{b}{Constrained by literature values where possible as described in Section \ref{sedmodeling}.}
\tablenotetext{c}{Fixed to 60$^{\circ}$ except for two objects with previous ALMA modeling. See Section \ref{sedmodeling}.}
\tablenotetext{d}{This is not a free parameter and was} calculated by DIAD.  See Section \ref{diskmasssection}
\tablenotetext{e}{Best fit as a pre-transitional disk. See Table \ref{table:ptd} for additional parameters.}
\tablenotetext{f}{Parameters from previous modeling by \citet{rilinger19}.}
\tablecomments{$\epsilon$, $\alpha$, a$_{max,atm}$, a$_{max,mid}$, H$_{in}$, and T$_{wall,in}$, in were varied to obtain the best fit starting with the initial grid in Table \ref{table:coarsegrid}.  The z$_{wall}$ values reported here are the H$_{in}$ values converted to units of au.}
\end{deluxetable*}

\subsubsection{Disk Mass}\label{diskmasssection}
\begin{figure*}
    \plotone{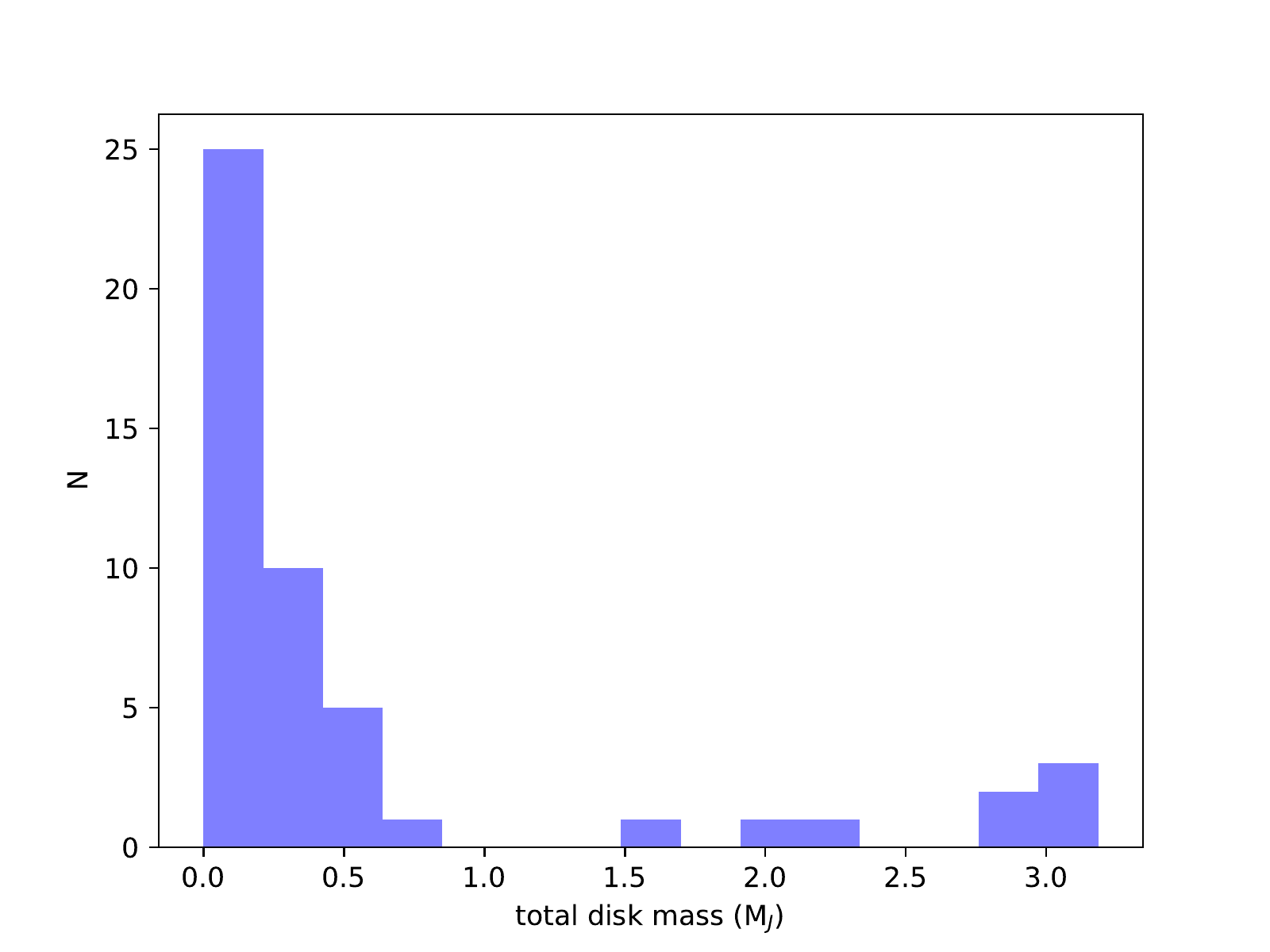}
    \caption{Histogram of the calculated disk masses for the BDs in this sample.  Only a few outliers have masses greater than 1 M$_J$, indicating that most of the BD disks are not currently massive enough to form planetary companions according to simulations by \citet{payne07}.}
    \label{fig:diskmass}
\end{figure*}

One of the most important disk properties to measure is the disk mass, since the mass of the disk constrains the potential for planetary companion formation.  As described above, the total (gas and dust) disk mass is determined by integrating over the surface mass density ($\Sigma$) calculated by DIAD.  For each BD in the sample, the disk mass was computed in this way using the $\Sigma$, $\alpha$, and $\dot{M}$ of the best-fit SED model.  A histogram of the disk masses found for this sample is shown in Figure \ref{fig:diskmass}.  The vast majority of disks in this sample have total disk masses below 1 M$_J$.  Seven disks (J04230607, KPNO Tau 3, J04400067, J04442713, CFHTWIR-Oph 66, GY92 264, and Lup 818s) are clear outliers, with more massive disks.

\subsubsection{Dust Growth}\label{growth}
Dust grains in the interstellar medium (ISM) are typically assumed to have a maximum size of 0.25 \mums \citep{mathis77, weingartner01}.  The presence of larger ($>$1 \mum) grains thus indicates that dust growth has taken place within the disk and therefore that the first steps towards planet formation have occurred.

The 10 \mums silicate feature commonly observed in TTS and BD disks contains significant information about grain growth in protoplanetary disks.  In these disks, this feature is often observed to be in emission \citep{cohen85, gurtler99, natta00}.  The presence of an emission line indicates that a (super)heated optically thin dust layer exists above the optically thick disk \citep{calvet91}.  In other words, the 10 \mums emission feature is produced by the disk atmosphere.  The shape of the feature corresponds to the size of grains in this atmosphere.  Submicron-sized grains show a more narrow, peaked shape; as grains grow to be larger than $\sim$ 1 \mum, the 10 \mums feature broadens and flattens.

The maximum size of dust grains in the disk atmosphere, a$_{max, atm}$, is one of the parameters varied in our SED models.  Furthermore, the wavelength coverage of IRS spectra encompasses the 10 \mums spectral feature.  Therefore, for the 34 objects in our sample with IRS spectra, we can use the model fit to the 10 \mums feature to probe grain growth.  Note that we do not attempt to fit the spectral feature in detail \citep[e.g.,][]{sargent09}; rather, we consider its width and height to generally constrain the grain size.  We assume the dust particles to be a standard mixture of olivine silicates and graphite.  Relative to the gas, silicates have a fractional mass abundance of 0.004 and graphite has a fractional mass abundance of 0.0025.  Much work has been done to understand how different dust compositions and opacities affect the SED \citep[e.g.,][]{espaillat10, espaillat12}; however varying the species and/or relative abundances is outside the scope of this work.  We assume the standard opacities calculated from Mie theory; optical constants are taken from \citet{dorschner95} and \citet{draine84} for the silicates and graphite, respectively.  

\begin{deluxetable}{c c c c c}
\tabletypesize{\footnotesize}
\tablecaption{Outer wall parameters\label{table:ptd}}
\tablehead{
\colhead{Object} & \colhead{z$_{wall, out}$} & \colhead{T$_{wall, out}$} & \colhead{R$_{wall, out}$} & \colhead{Gap width}\\  & au & K & au & au
}
\startdata
J04442713 & 0.0827 & 400 & 0.27 & 0.20\\
CFHT Tau 9 & 0.0275 & 300 & 0.36 & 0.34\\
\enddata
\end{deluxetable}

\begin{figure}
    \plotone{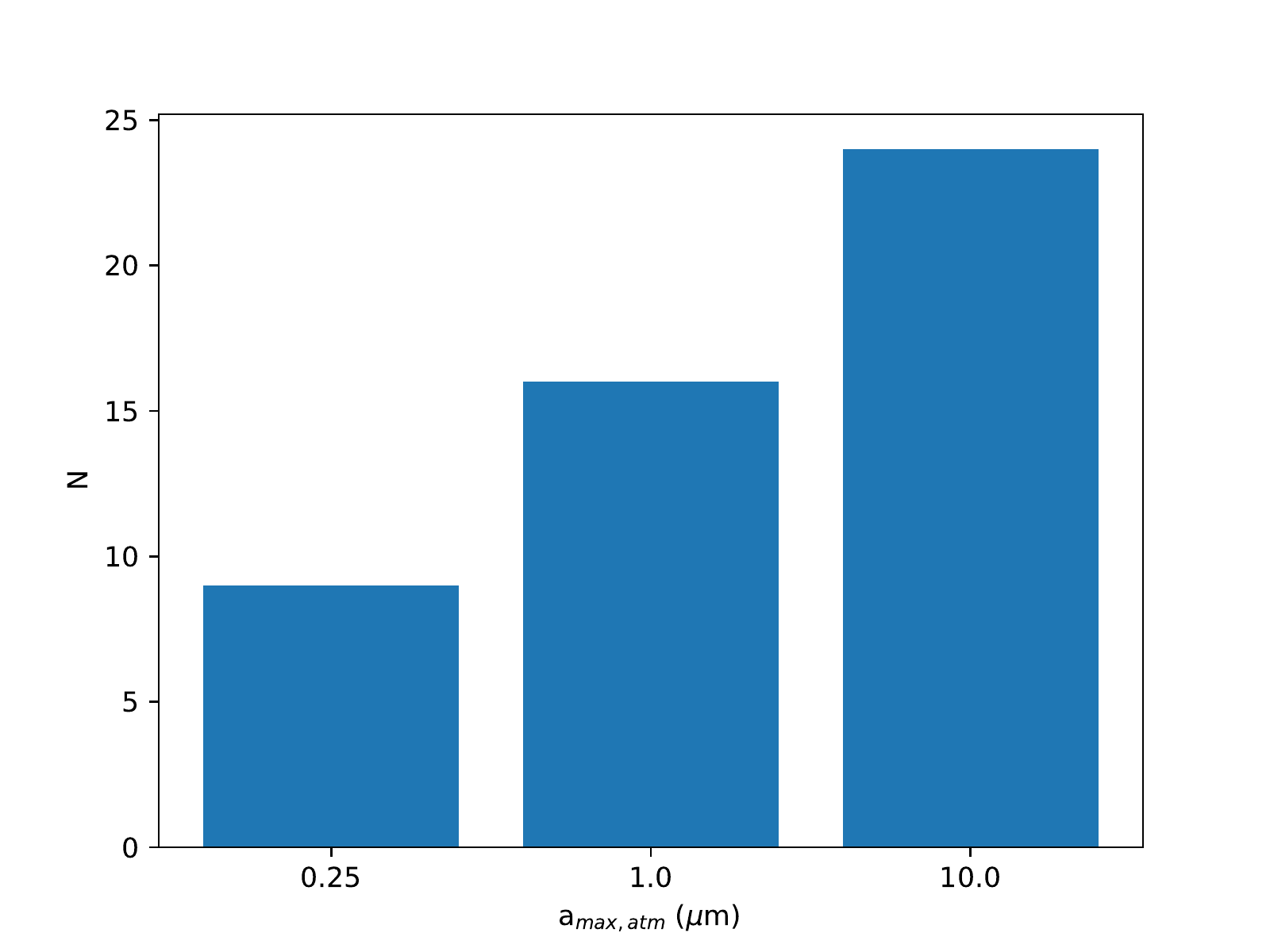}
    \caption{Histogram of the best-fit a$_{max, atm}$ values for the BDs in this sample.  The majority of disks in this sample have a$_{max, atm}$ greater than 0.25 \mum, indicating grain growth in these disks.}
    \label{fig:amax}
\end{figure}

We ran models with three different a$_{max, atm}$ for atmospheric dust grains: 0.25 \mums (i.e., ISM grains), 1.0 \mums (moderate grain growth), and 10.0 \mums (significant grain growth).
Figure \ref{fig:amax} shows a histogram of the a$_{max, atm}$ values from the best-fit models to each BD SED.  The majority of disks are best fit with grains larger than standard ISM grains, indicating that significant grain growth has taken place in these disks.

\subsubsection{Dust Settling}\label{settling}
Another expected process in disk evolution is the settling of dust grains towards the disk midplane as the grains grow larger.  To parameterize the amount of dust settling, we assume the grain size distributions and the dust-to-gas mass ratio ($\zeta$) to be constant in radius and vary only with height above and below the disk midplane.  In order to keep the total $\zeta$ constant while still allowing for settling of large grains, we decrease the dust-to-gas ratio of small grains ($\zeta_{small}$) in the atmosphere and increase the dust-to-gas ratio of larger grains ($\zeta_{big}$) in the midplane. We thus define the settling parameter $\epsilon$ such that
\begin{equation}
    \epsilon = \zeta_{small} / \zeta_{std}
\end{equation}
where $\zeta_{std}$ is the standard assumed dust-to-gas mass ratio of 0.01.  Therefore, lower $\epsilon$ values correspond to more settled disks.

\begin{figure}
    \plotone{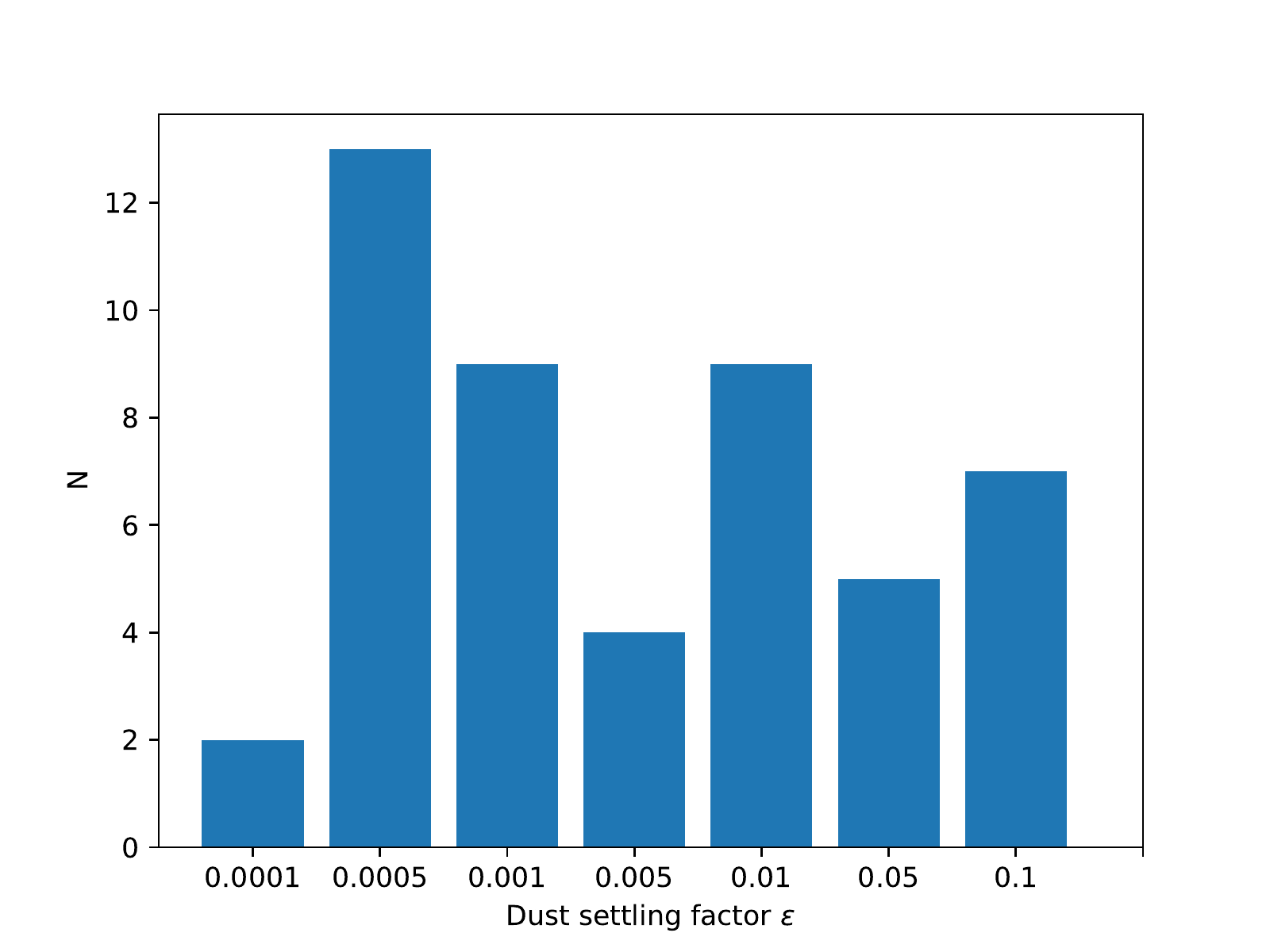}
    \caption{Distribution of the dust settling parameter $\epsilon$ for all BDs in this sample.}
    \label{fig:epshist}
\end{figure}

In our grid of models, we vary $\epsilon$ between 0.0005 and 0.1.  The histogram of best-fit $\epsilon$ values for the disks in this sample is shown in Figure \ref{fig:epshist}.  The distribution peaks towards lower $\epsilon$ values, indicating a high degree of settling in BD disks.  In conjunction with the grain growth present in many BD disks modeled here, this high degree of settling indicates that BD disks show significant evidence of dust evolution.

\subsubsection{Disk Clearing}
Disks that show evidence of clearing (i.e., pre-transitional and transitional disks; PTDs and TDs, respectively) are not uncommon around around T Tauri stars \citep[e.g.,][]{espaillat07, muzerolle10, kim13}.  BD disks may also exhibit similar substructures.  

In young stellar objects (YSOs), dust sublimation temperatures typically range between 1000--2000 K \citep{monnier02}.  Therefore, inner wall temperatures below 1000 K indicate additional disk clearing beyond the dust sublimation radius expected in the inner disk.  In other words, inner wall temperatures below 1000 K correspond to a larger inner disk radius and thus a dust-depleted hole in the disk.  In our sample of BDs, 7 disks were best fit with inner wall temperatures below 1000 K and are therefore classified as TDs (see Table \ref{table:params}, column 7).

Additionally, two objects in this sample were best fit as PTDs, having an inner disk and outer disk separated by a dust-depleted gap.  These objects and their outer disk parameters are listed in Table \ref{table:ptd}.  One of these PTDs, J04442713, was previously reported by \citet{rilinger19}.  We find that 9 objects, 18 percent of the disks studied here, show some evidence of disk clearing.  This fraction of disks with evidence of clearing is consistent with the fraction of TDs that has been reported for TTS \citep[between 10 and 20 percent of objects, see][]{muzerolle10}.

It is important to note that these holes and gaps were inferred from the SEDs of these objects and have not been directly detected through mm-wavelength imaging (e.g., with ALMA).  Most of the TDs presented here do not have clear dips in their SEDs as are seen in TTS TDs.    \citet{ercolano09} note that the unambiguous detection of TDs around late-M stars is complicated by their lower stellar luminosities: the ratio between the dust and stellar contributions to the SED is smaller (i.e., the contrast is lower) than for earlier-type stars.  This challenge in interpreting the SED means that high-resolution imaging of these objects is necessary to confirm the presence of these substructures.  However, given the small inner hole radii and gap widths (see Tables \ref{table:params} and \ref{table:ptd}), these features may be below current resolution limitations.

\section{Discussion}\label{discussion}
Having obtained models for all 49 objects in our sample, we can compare how the four star-forming regions differ from one another.  Since the four regions studied here vary in age, differences between regions correspond to how disks evolve over time.  We also consider how the BD disks differ from previously-studied TTS disks to determine what affect host mass has on certain disk properties.  Disk mass is of particular interest; we assess the implications the BD disk masses have for planet formation and measure the relationship between the host masses and disk masses.

\subsection{Comparison Between Star-Forming Regions}\label{comparison}
\subsubsection{Disk Mass}\label{diskmasssubsection}
\begin{figure*}
    \plotone{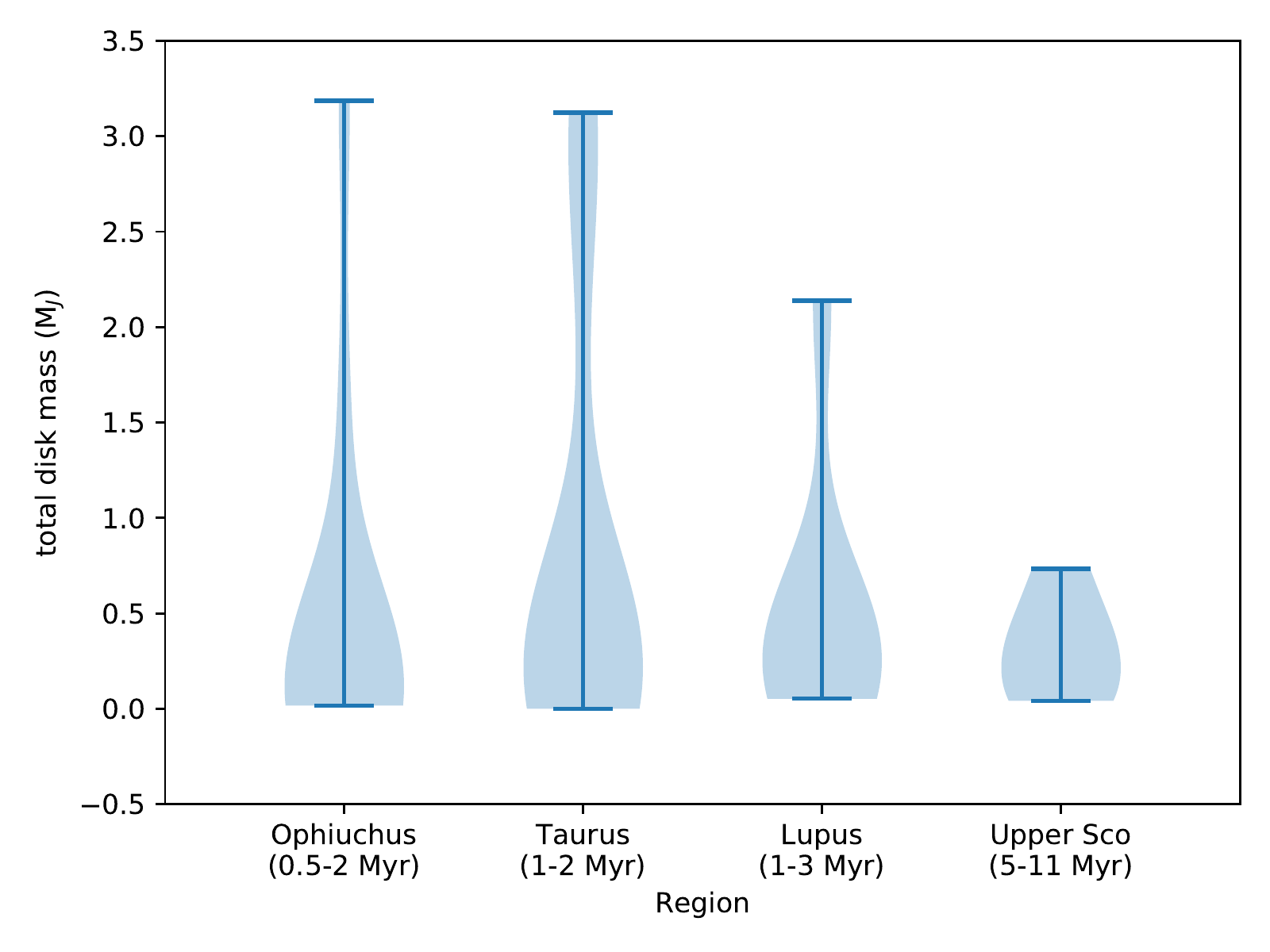}
    \caption{Calculated disk masses for all BDs in this sample, separated by region.  The four regions are ordered by age, increasing left to right.  The blue lines show the total range of disk masses in each region; the width of the shaded blue regions represent the distributions of the disk mass values.}
    \label{fig:masscomp}
\end{figure*}

Average disk masses for TTS have been reported to decrease with age of the star-forming region \citep{ansdell16, barenfeld16, pascucci16, cieza19, vanterwisga19}.  Figure \ref{fig:masscomp} shows how disk mass varies with age for BD disks to determine whether this trend holds for substellar objects as well.  Two main deviations from the trend of decreasing disk mass with age can be seen: (1) Upper Sco, the oldest region in the sample, has only slightly lower disk masses than the other regions and (2) Ophiuchus, the youngest region in the sample, does not have significantly higher disk masses than the older regions.

The unexpectedly high mass of the Upper Sco disks is likely due to selection effects and the substellar masses of the host BDs.  In order to maximize the likelihood of detecting disk emission in their ALMA survey of BDs and low mass stars in Upper Sco, \citet{vanderplas16} only observed objects with the largest mid-IR excesses and required the objects to have detections in both the \textit{Herschel} PACS 70 \mum{} and 160 \mum{} channels.  Therefore, the Upper Sco disks considered here do not represent a complete sample of BD disks in the region; moreover, the disks that did not meet the mid-IR excess or PACS detection requirements are likely less massive than the disks included in the sample.  

However, \citet{carpenter06}, \citet{bayo12}, and \citet{luhmanmamajek12} find evidence that disk lifetimes are longer for lower mass hosts.  Though Upper Sco is older than the other regions studied here, we would still expect to see massive disks around BDs in that region if their disks dissipate more slowly than their TTS counterparts.

\begin{figure}
    \plotone{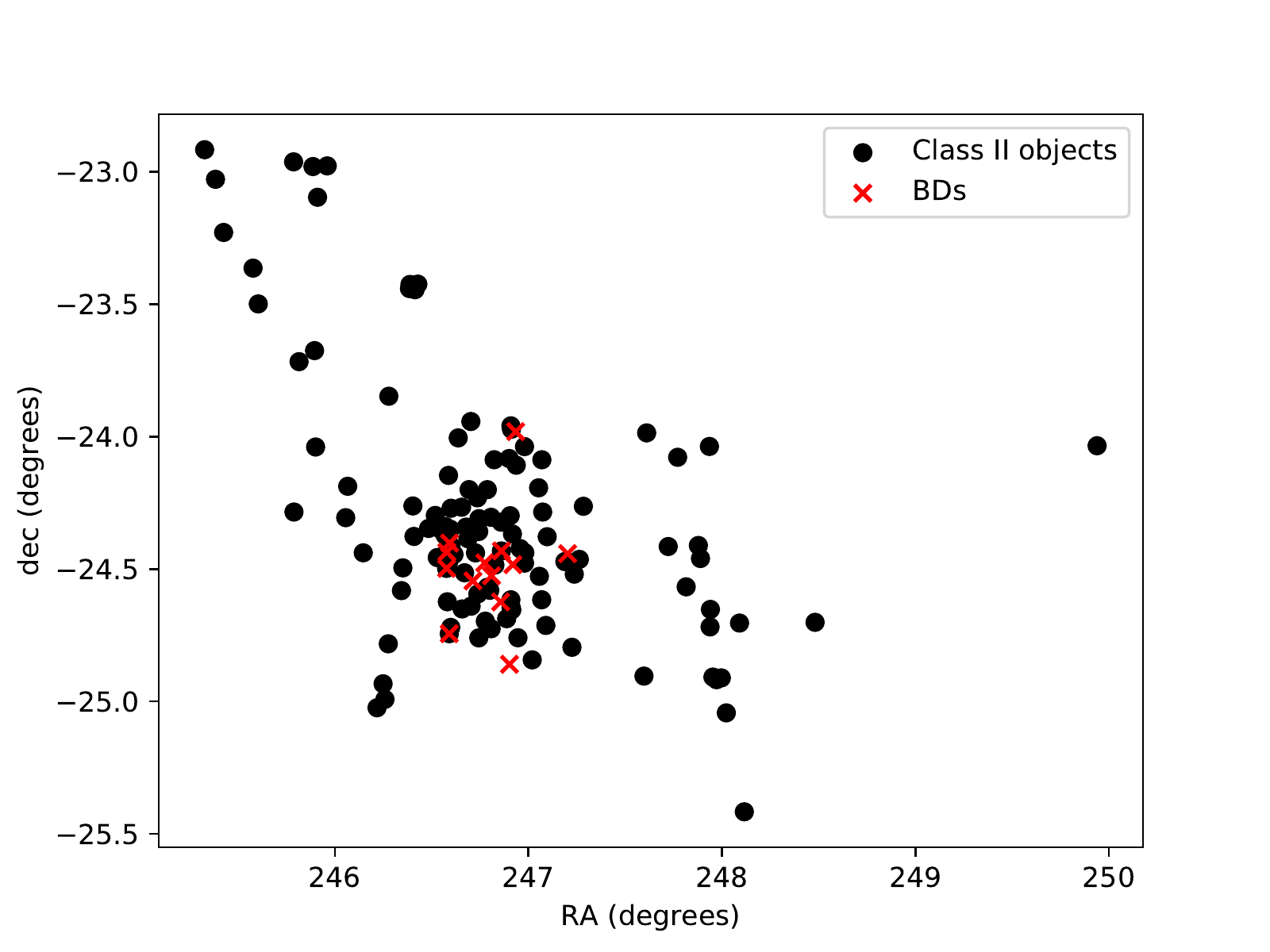}
    \caption{Location of the TTS studied by \citet{williams19} (black circles) and the BD disks observed by \citet{testi16} and Testi et al., submitted and studied in this work (red crosses).  Any local cloud environment effects that influence the disk size of the TTS would affect the BDs as well since their positions in space overlap.}
    \label{fig:oph_locs}
\end{figure}

As the youngest region in the sample, disks in Ophiuchus should be the most massive.  However, we find that the disk masses in Ophiuchus are relatively comparable to the older Taurus and Lupus regions.  All three regions have at least one disk greater than 1 M$_J$.  In fact, Taurus has a slightly greater fraction of massive disks than Ophiuchus. \citet{williams19} also found TTS disks in Ophiuchus to be less massive than expected.  In their study, they find the Ophiuchus disks to be slightly less massive than TTS disks in the somewhat older Lupus region.  These authors suggest the local cloud environment could have a substantial effect on the disk masses in a given region.  For example, \citet{kuffmeier20} find that ionization can decrease disk size: disks may be ``born'' with smaller masses in star-forming regions with higher ionization rates.  Figure \ref{fig:oph_locs} plots the locations of the Ophiuchus BDs in this sample along with the locations of the TTS disks from \citet{williams19}.  The two populations overlap, so it is reasonable to assume that any local cloud environment effects that influence the disk masses of the TTS would also influence the disk masses of the BDs.

Of the four star-forming regions studied here, Taurus contains the most massive disks, including four disks with masses greater than 1 M$_{Jup}$ (KPNO Tau 3, J04400067, J04442713, and J04230607), though Lupus and Ophiuchus both contain at least one equally massive disk (Lup 818s in Lupus, and CFHTWIR-Oph 66 and GY92 264 in Ophiuchus).  The high mass of J04442713 was previously reported by \citet{rilinger19}; the finding of three additional objects in the same region with similar masses may indicate that there is a population of high-mass BD disks in Taurus.  As the youngest region at $\sim$0.5 Myr, disks in Ophiuchus are expected to have not yet dissipated. At $\sim$1 Myr, Taurus is also a relatively young star-forming region; Lupus is slightly older, at $\sim$1--3 Myr, but still young enough that significant disk dissipation may not yet be common.

\begin{figure*}
    \gridline{\fig{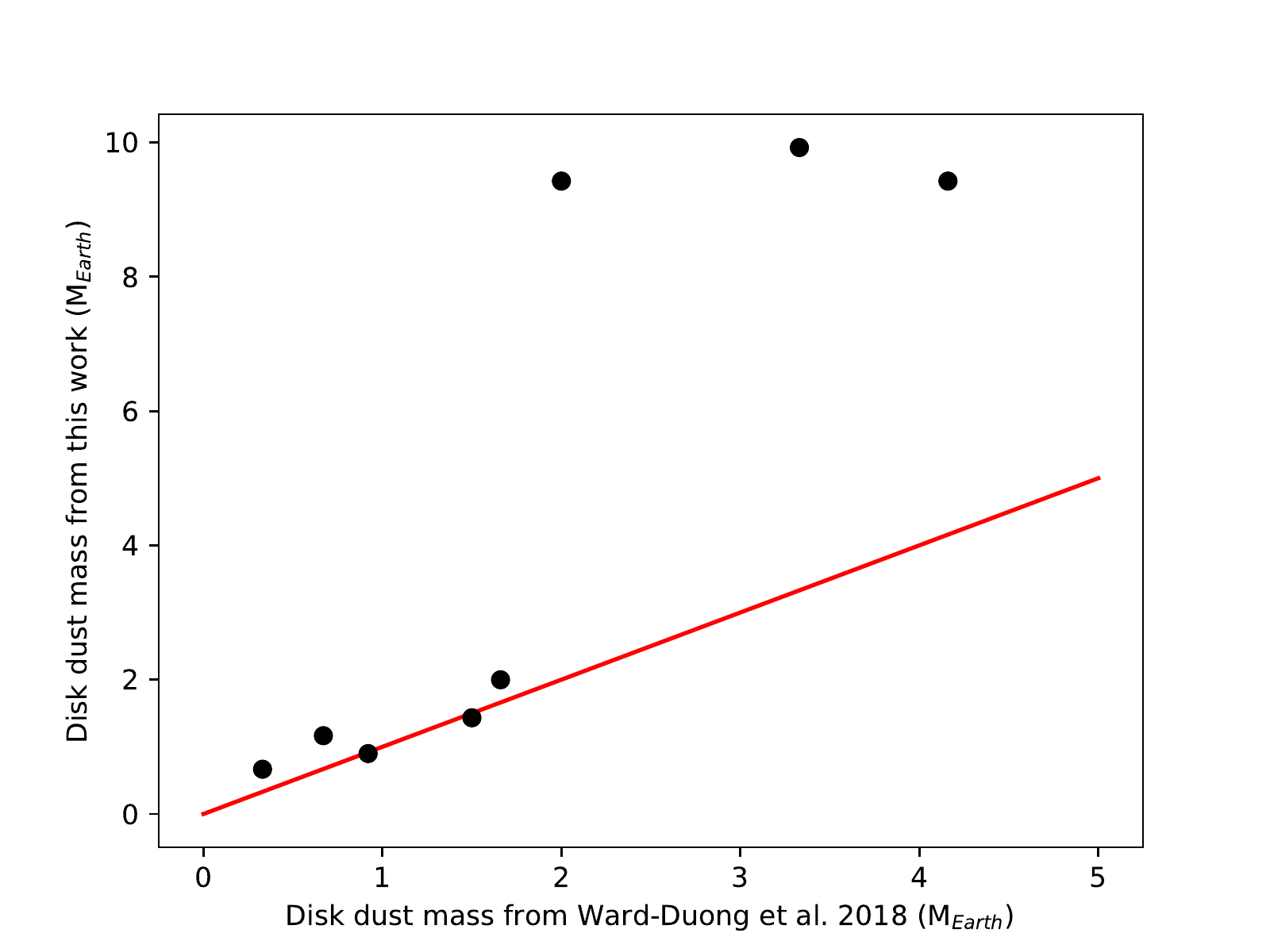}{0.5\textwidth}{(a)}
            \fig{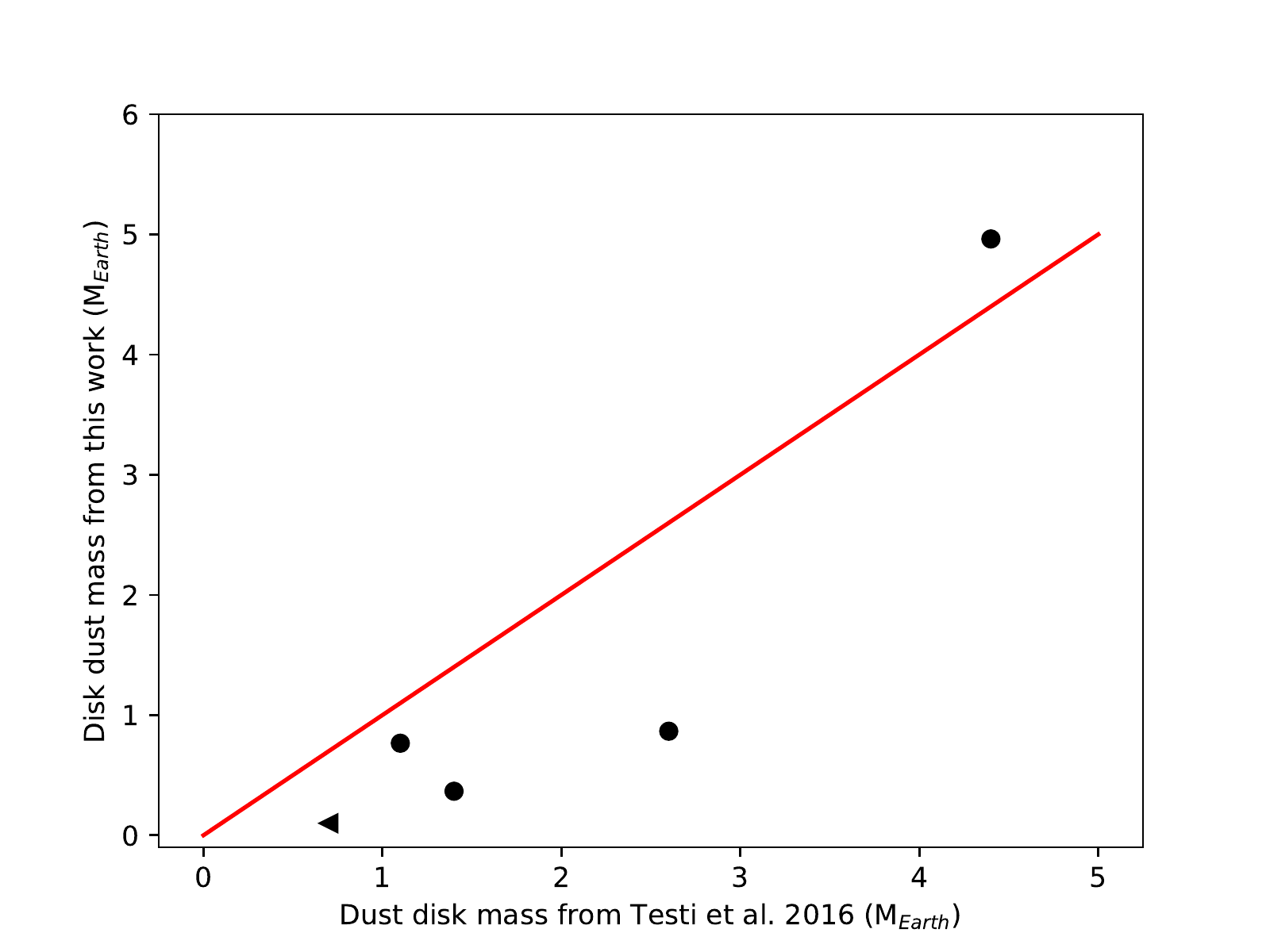}{0.5\textwidth}{(b)}}
    \gridline{\fig{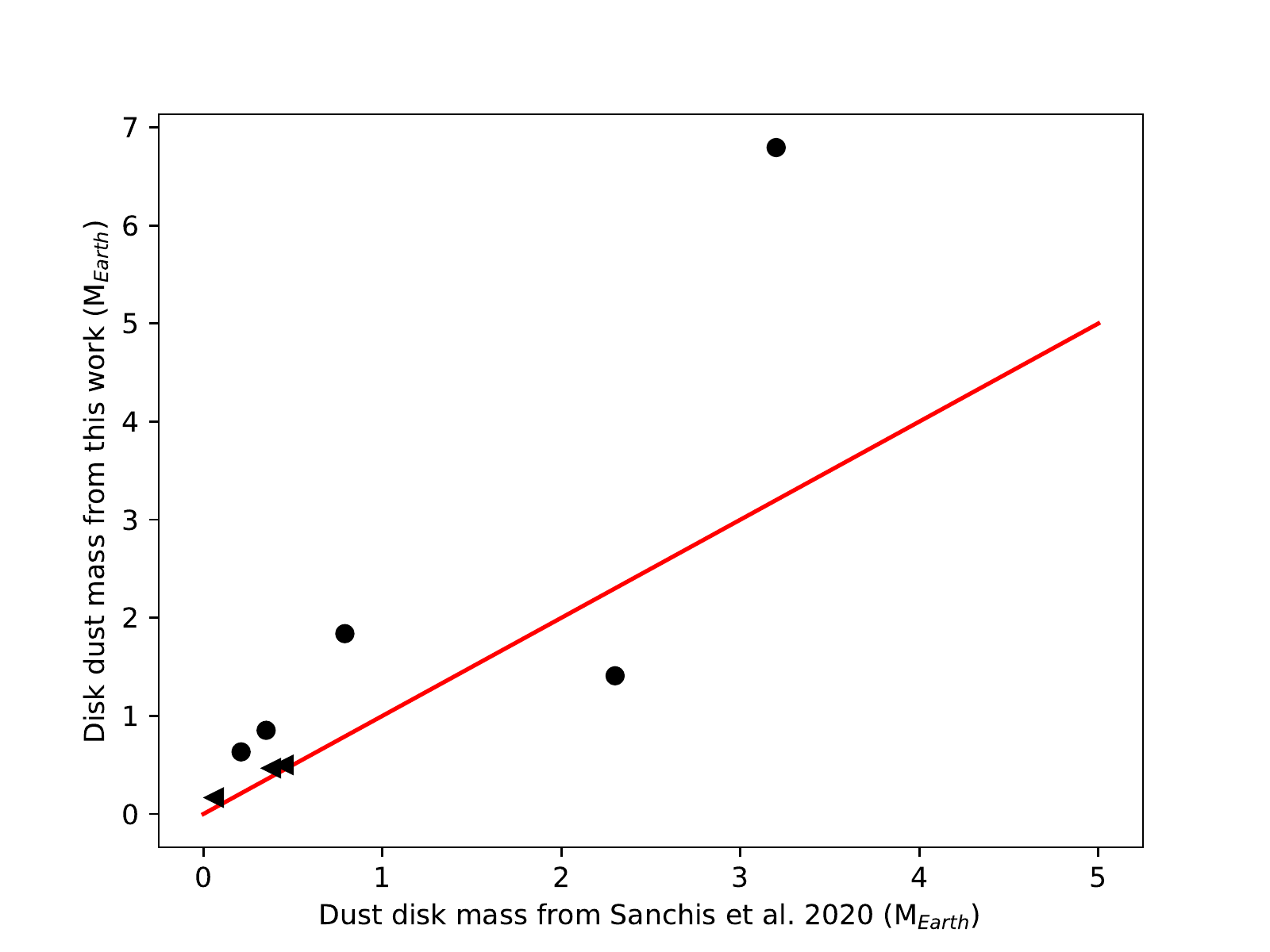}{0.5\textwidth}{(c)}
            \fig{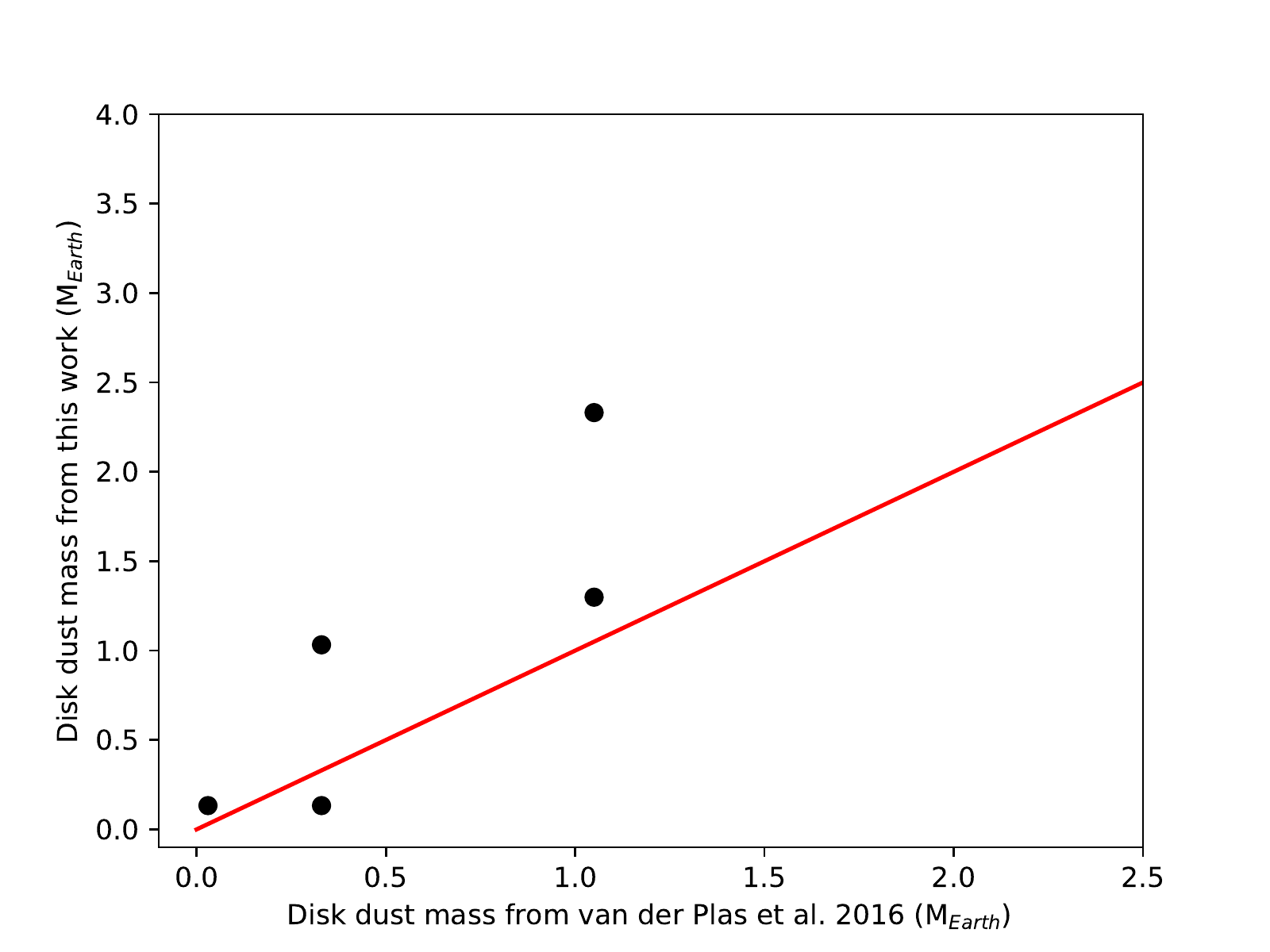}{0.5\textwidth}{(d)}}
    \caption{Comparison of disk dust masses calculated in this work to those reported by (a) \citet{ward-duong18}, (b) \citet{testi16}, (c) \citet{sanchis20}, and (d) \citet{vanderplas16}.  In each panel the red line indicates a 1:1 relationship.  Triangles indicate upper limits.}
    \label{fig:refmasscomp}
\end{figure*}

Figure \ref{fig:refmasscomp} compares the dust disk masses calculated in this work to previously reported masses in the literature.  In particular, we compare our results to \citet{ward-duong18}, \citet{testi16}, \citet{sanchis20}, and \citet{vanderplas16} for objects in Taurus, Ophiuchus, Lupus, and Upper Sco, respectively.  \citet{ward-duong18} and \citet{vanderplas16} also use SED fitting to determine disk masses for their sample of BDs.  Both works use the MCFOST radiative transfer code \citep{pinte06, pinte09}.  As shown in the panels (a) and (d) of Figure \ref{fig:refmasscomp}, we determine disk masses that are generally in good agreement, though we find three disks (J04230607, KPNO3, and J04400067) to be $\sim$2-3 times more massive (note that \citet{ward-duong18} did not study the fourth massive disk in Taurus, J04442713).  None of these three massive disks have measured $\dot{M}$ values.  As discussed in Section \ref{diskmasssection}, $\dot{M}$ directly impacts our disk mass calculation; thus, if the $\dot{M}$ values for these objects are significantly lower than the assumed value of 10$^{-10}$ M$_{\odot}$ yr$^{-1}$, these disk masses would be overestimated.  Overestimating the radii of these disks could also result in an overestimate of the disk mass, since we integrate the surface mass density over the disk radius.  High-resolution ALMA images of these disks would help constrain their radii and thus the disk mass.  

We note that the fourth massive disk reported here, J04442713, was determined by \citet{rilinger19} to have a large disk radius of 100 au based on ALMA observations, so its high mass is consistent with the disk being large.  This object also has a measured accretion rate of 2 x 10$^{-10}$ M$_{\odot}$ yr$^{-1}$, so its disk mass is more certain than the other three massive disks.

\citet{testi16} and \citet{sanchis20} do not use SED fitting to determine disk mass; rather, they use measurements of the mm flux (F$_{\nu}$) of each BD to calculate the disk mass according to the following relationship:
\begin{equation}\label{dustmass_conversion}
   M_{dust} = \frac{F_\nu d^2}{\kappa_\nu B_\nu(T)} 
\end{equation}
where d is the distance to the BD, $\kappa_{\nu}$ is the opacity at frequency $\nu$, and B$_\nu$(T) is the Planck function at temperature (\textit{T}).  This relationship assumes the emission is optically thin and isothermal at \textit{T}.  For most objects in the \citet{testi16} sample, we find disks that are less massive than their reported values (see Figure \ref{fig:refmasscomp}, panel (b)).  This discrepancy could indicate that either or both of the assumptions of optically thin and/or isothermal emission are incorrect.  The DIAD model does not assume an isothermal disk, but rather calculates the temperature profile of the disk.  Using radiative transfer models of protoplanetary disks in Taurus, \citet{ballering19} find that the optically thin assumption frequently does not hold and that the disks tend to be somewhat optically thick.  This conclusion is also supported by neural network-based modeling with DIAD performed by \citet{ribas20}.  Thus, the assumption of optically thin emission in Equation \ref{dustmass_conversion} is likely the source of the difference with the \citet{testi16} disk mass estimates.  However, the results from \citet{testi16} do agree that GY92 264 is a substantially more massive disk than the other disks in Ophiuchus.  We note that \citet{testi16} did not study CFHTWIR-Oph 66, the other massive Ophiuchus disk in our sample.

The masses we determine for the Lupus objects are in somewhat better agreement with the \citet{sanchis20} masses, though our mass for Lup 818s is about 2 times more massive than they report.  Unlike J04230607, KPNO3, and J04400067 discussed above, Lup 818s has a measured accretion rate, so uncertainty in $\dot{M}$ is likely not the cause for the large mass.  Furthermore, \citet{sanchis20} determined an upper limit to the disk radius of $\sim$20 au for this object from their ALMA observations.  This compact yet massive disk may be an interesting object for follow-up ALMA observations.

\subsubsection{Dust Growth}
\begin{figure*}
    \plotone{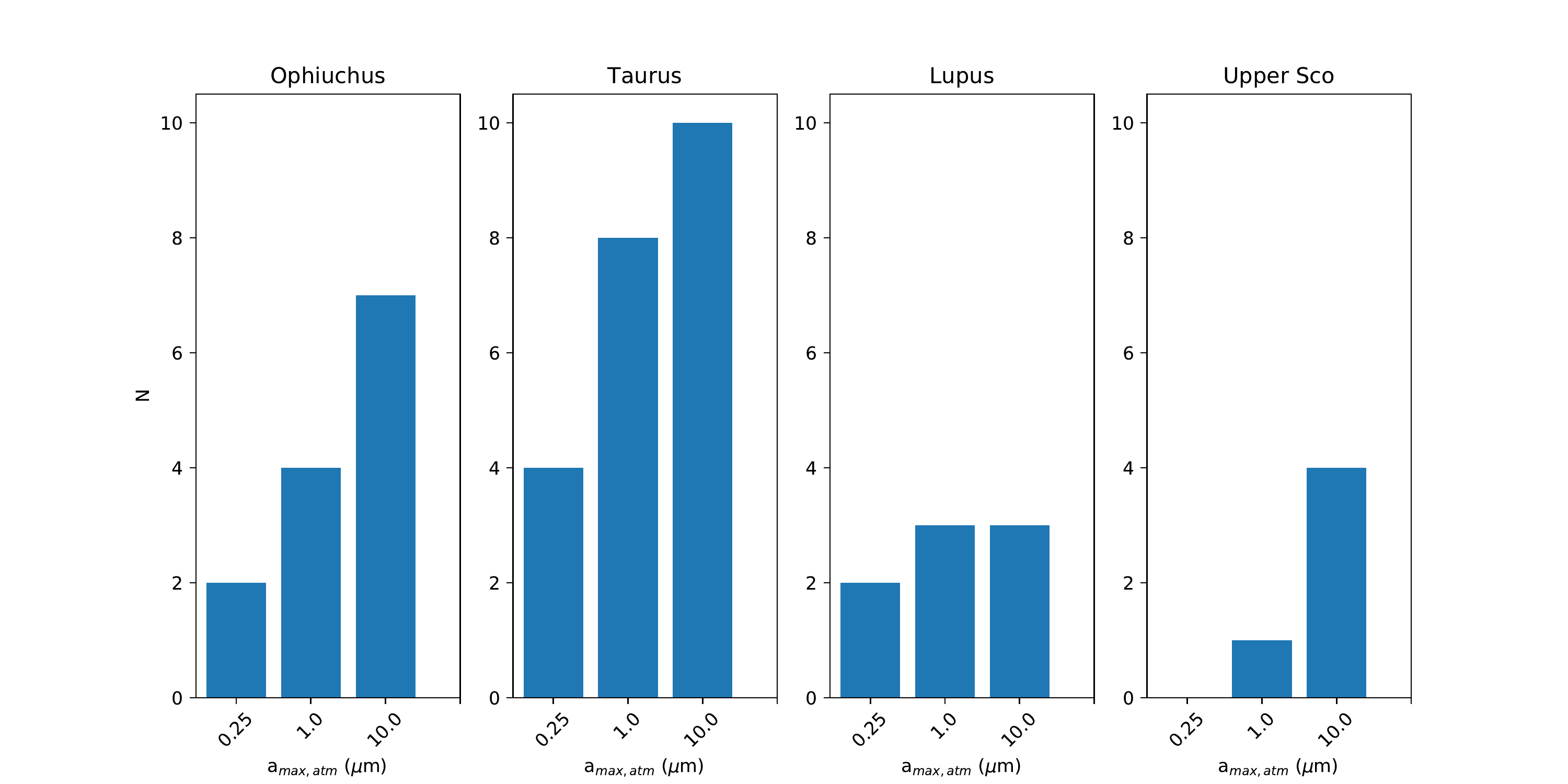}
    \caption{Distribution of the maximum size of dust grains in the disk atmosphere, a$_{max, atm}$, for all BDs in this sample, separated by region.  The four regions are ordered by age, increasing left to right.}
    \label{fig:amaxcomp}
\end{figure*}

As discussed in Section \ref{growth}, dust grains are expected to increase in size as disks evolve.  Figure \ref{fig:amaxcomp} shows the distribution of the best-fit atmospheric dust grain maximum sizes from our models for the BD disks in each of the four star-forming regions studied here.  The histograms show that all regions, even the youngest, have undergone some amount of grain growth.  All disks in Upper Sco, the oldest region, have grains larger than typical ISM grains; in fact all but one of the disks in this region have grains about two orders of magnitude larger than ISM-sized grains.  

The result that the oldest star-forming region in our sample has the largest dust grains is consistent with the idea that the oldest disks should be the most evolved.  It is important to note, however, that all four regions contain at least some disks with evidence of grain growth.  Grain growth occuring even in young star-forming regions is consistent with previous studies of TTS disks in Taurus, Ophiuchus, Lupus, and Chamaeleon, which report grain growth to millimeter sizes in the disk midplanes based on spectral indices \citep[e.g.,][]{lommen07, ricci10a, ricci10b, ribas17}.  Grain growth at young ages supports the hypothesis that planets may form quickly \citep[e.g.,][see Section \ref{massdiscussion} of this work.]{manara18}.

\subsubsection{Dust Settling}
\begin{figure*}
    \plotone{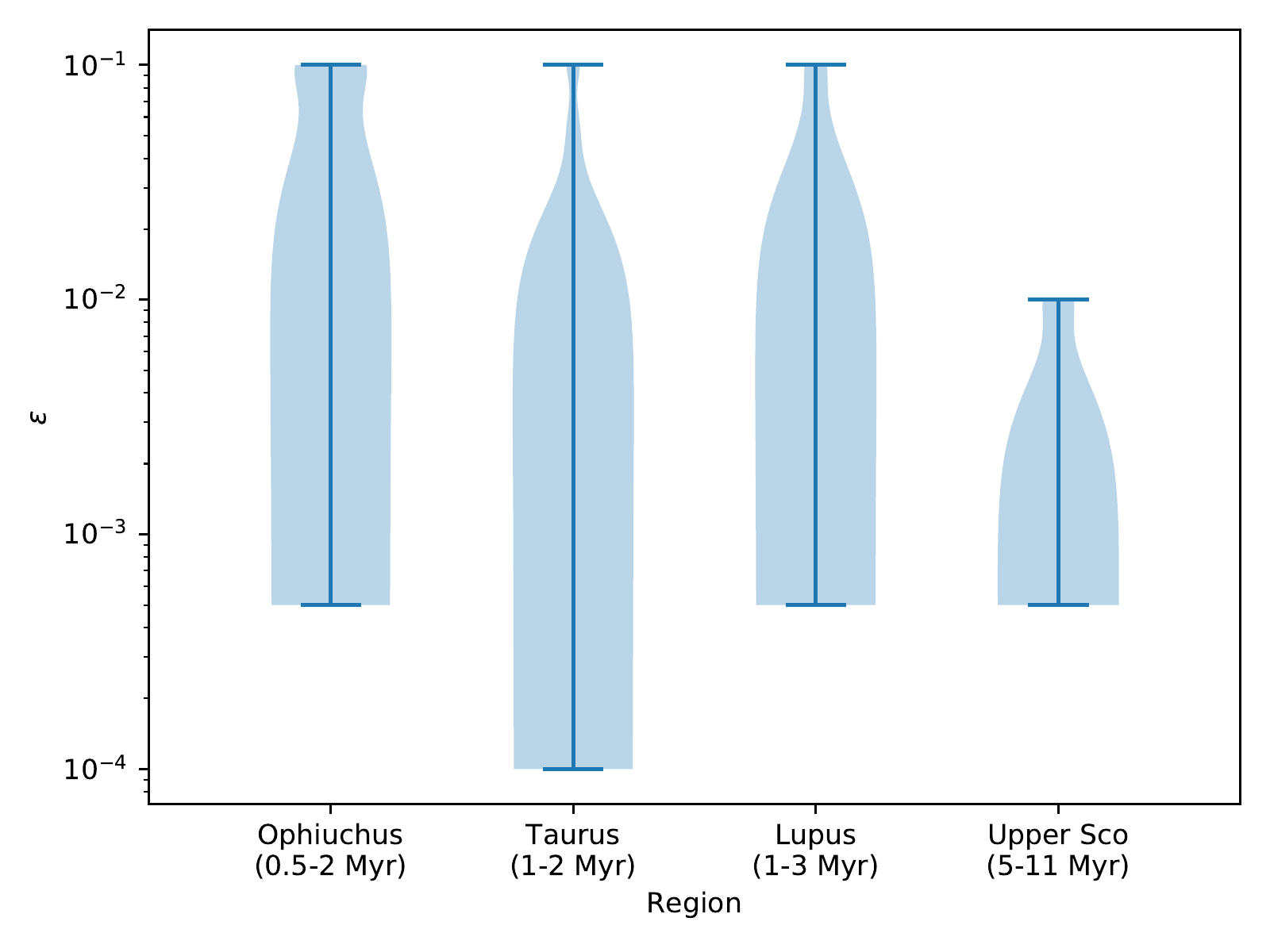}
    \caption{Distribution of the dust settling parameter $\epsilon$ for all BDs in this sample, separated by region.  The four regions are ordered by age, increasing left to right.}
    \label{fig:epscompviolin}
\end{figure*}

Figure \ref{fig:epscompviolin} shows the distribution of the best-fit settling parameter $\epsilon$ from our models for the BD disks in each of the four star-forming regions studied here.  Lower values of $\epsilon$ correspond to higher degrees of settling.  We expect the degree of settling to increase with age and amount of grain growth.  Indeed, Ophiuchus has the largest fraction of high $\epsilon$ values (i.e., the least settled disks) and Upper Sco shows the lowest average $\epsilon$ value in its disks, which is consistent with its larger average grain size and older age.  However, the three younger regions contain some settled disks as well, though they also have less-settled disks than Upper Sco.  Evidence of settling in young disks has previously been reported by \citet{furlan09}, \citet{lewis16}, and \citet{grant18}, among others.  Along with evidence of grain growth (see previous section) and disk substructures \citep[e.g., HL Tau][]{alma15} in young ($\sim$1 Myr old or younger) objects, these low $\epsilon$ values are consistent with disk evolution occurring sooner than previously thought.  

\begin{figure*}
    \plotone{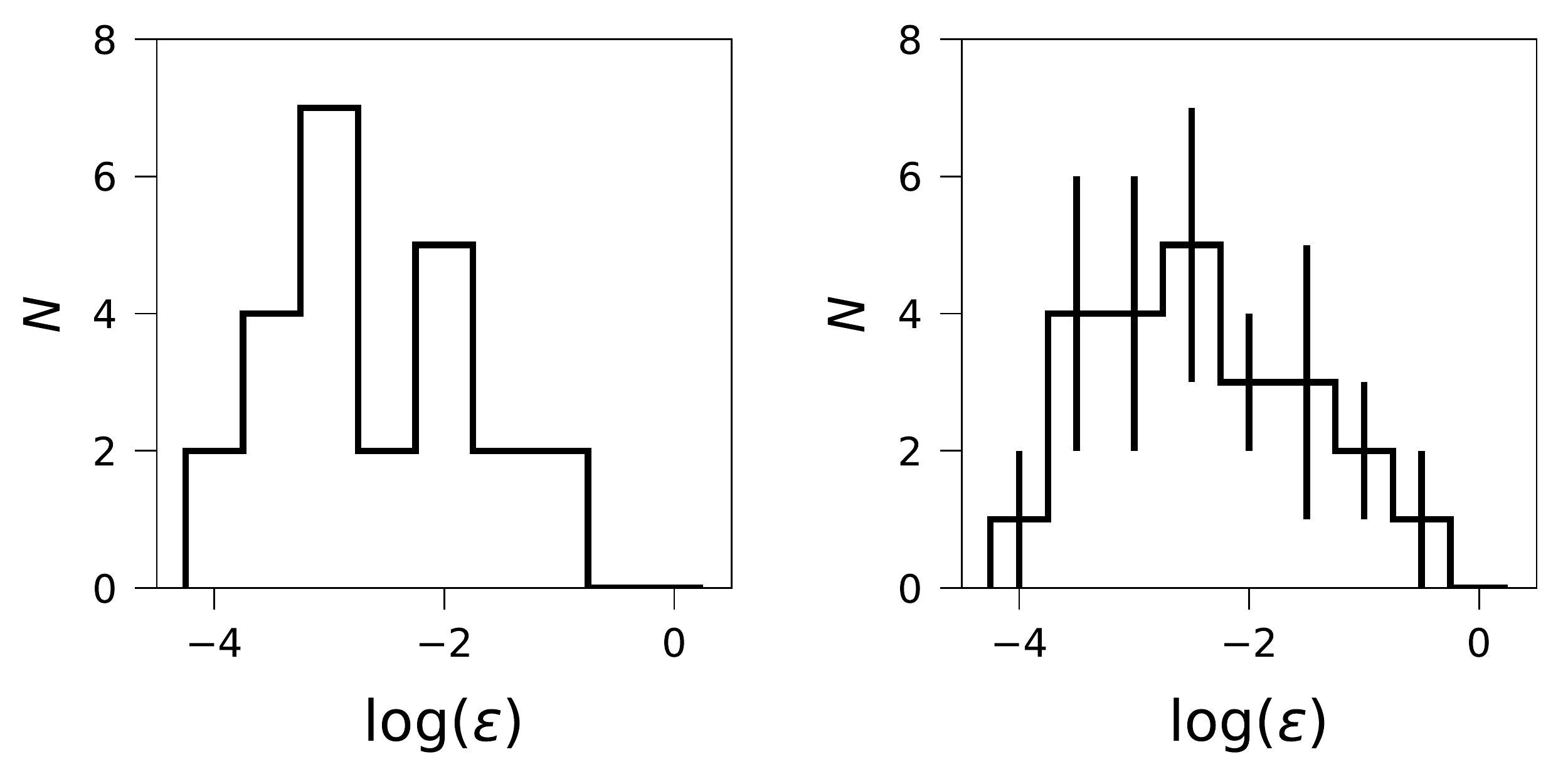}
    \caption{Distribution of the dust settling parameter $\epsilon$ for BDs (left) and TTS (right) in Taurus.  Right panel adapted from \citet{grant18}.  Error bars are standard deviations of the forward-modeling realizations performed by \citet{grant18}.}
    \label{fig:epscomp}
\end{figure*}

Figure \ref{fig:epscomp} compares the settling measured for Taurus BDs in this sample to the settling of TTS disks in Taurus as modeled by \citet{grant18}.  \citet{grant18} used a forward modeling process and DIAD to predict the distribution of $\epsilon$ for TTS disks in the Taurus Molecular Cloud.  This distribution is not a histogram of the $\epsilon$ values from best-fit SEDs for individual objects.  However, it is still useful to compare this distribution to our histogram of best-fit $\epsilon$ values for BDs in Taurus in order to search for similarities or differences in the amount of settling for these two types of objects.

For both BDs and TTS in Taurus, the distribution of $\epsilon$ values peaks at lower epsilon values.  Thus, both populations of objects appear to display similarly high degrees of settling.  BDs appear to evolve similarly to TTS over time, with perhaps longer lifetimes as reported by \citet{luhmanmamajek12}.

\subsection{Disk Mass and Planet Formation}\label{massdiscussion}
The amount of mass in a disk is directly linked to the potential for planet formation.  In their simulations of planet formation in the disks around BDs, \citet{payne07} found that Earth-mass planets are able to form easily in disks of $\sim$1 Jupiter mass or more, but lower mass disks were very unlikely to form planets of any mass.  These authors ascribe the lack of planets in lower mass disks to the lower disk surface density.  Taking the 1 M$_J$ lower limit as a ``cutoff'' for planet formation, very few disks in our sample are likely to form Earth-mass planets and the possibility of planets forming in the bulk of BD disks is therefore slim.  Only seven disks, discussed in Section \ref{comparison}, have masses above this cutoff.

As shown in Figure \ref{fig:refmasscomp}, previous works also report small disk masses for BD disks.  Furthermore, studies of protoplanetary disks around TTS, e.g., \citet{najita14}, find masses too small to account for observed planetary systems. However, though the current masses of these disks are too small for planet formation, our models do not rule out the possibility that planets may have already formed in the disks.  \citet{manara18} suggested rapid planet formation within $<$0.1--1 Myr could explain the discrepancy between disk mass and observed planetary systems.  Observations of disk substructures in 0.1--1 Myr protostellar disks \citep{sheehan20} as well as grain growth and settling in $\sim$1 Myr-old disks \citep{grant18} are evidence of early-onset disk evolution and possibly planet formation.

Alternatively, if grain growth in BD disks is efficient (i.e., grains stick together upon collision without splintering, and photoevaporation is inefficient), a significant amount of the disk material could be composed of bodies larger than $\sim$1 m but smaller than planets.  These bodies would not contribute to the emissivity at millimeter wavelengths, and thus would not have an effect on the SEDs of the disks.  Yet another possibility is that disks are replenished over their lifetimes by material from the surrounding interstellar medium \citep{manara18}, so that the total amount of planet-forming material available is greater than the amount of material in the disk at any given time.  However, the exact mechanism by which planets form in these disks is still unknown, and the origin of observed planetary-mass companions to BDs remains mysterious.

\subsection{Disk Mass -- Host Mass Relationship}
Sub-millimeter surveys of disks in various star-forming regions have established a positive correlation between the masses of the disks and the masses of the stars that host them \citep[e.g.,][]{andrews13, ansdell16, barenfeld16, pascucci16}.  For the most part, these surveys have focused on stellar-mass hosts and have not included many substellar-mass (i.e., BD) hosts.  In this section, we explore how this mass relationship changes when BD disks are included in the sample.

\begin{figure*}
    \gridline{\leftfig{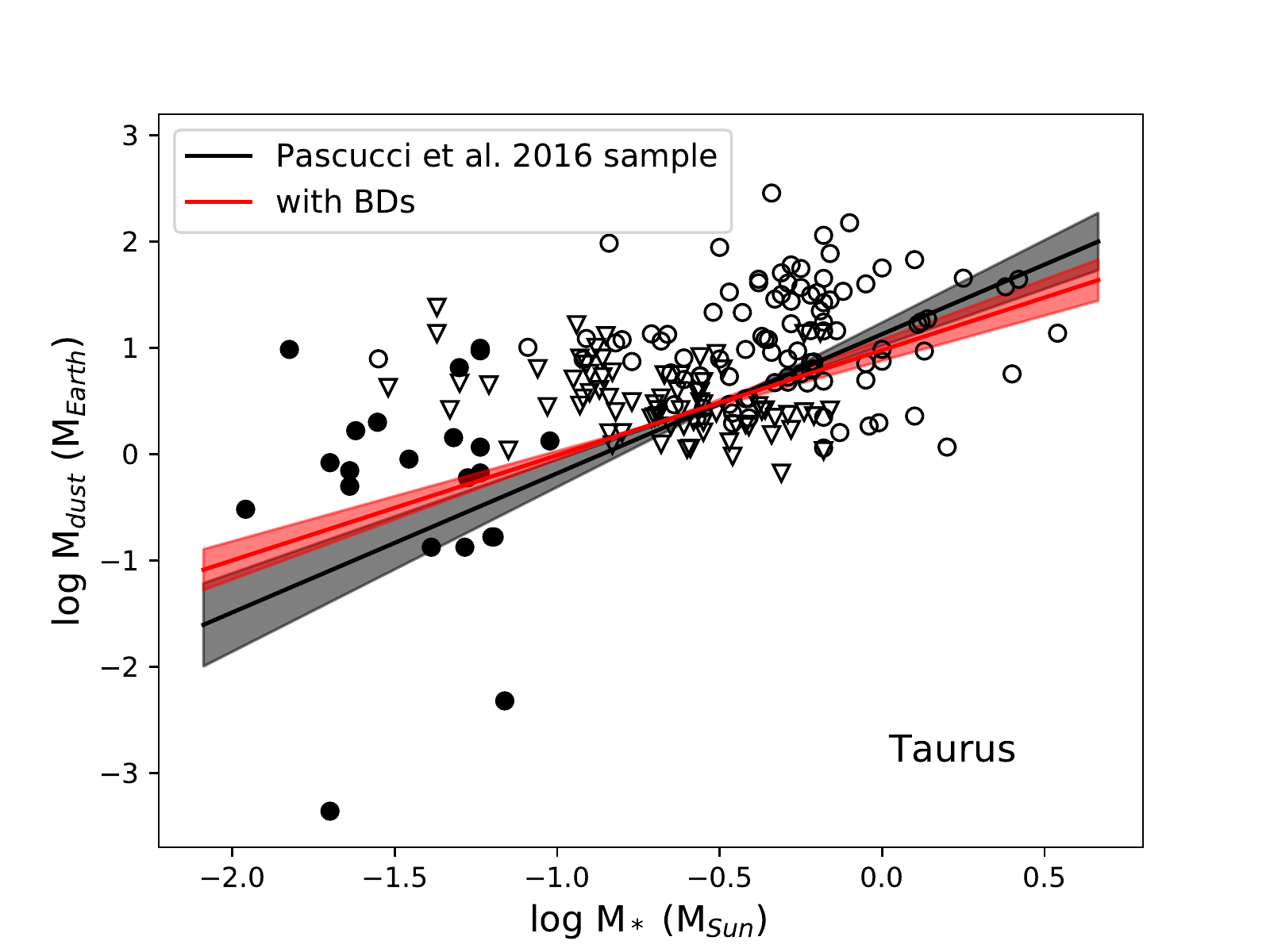}{0.34\textwidth}{(a)}
            \fig{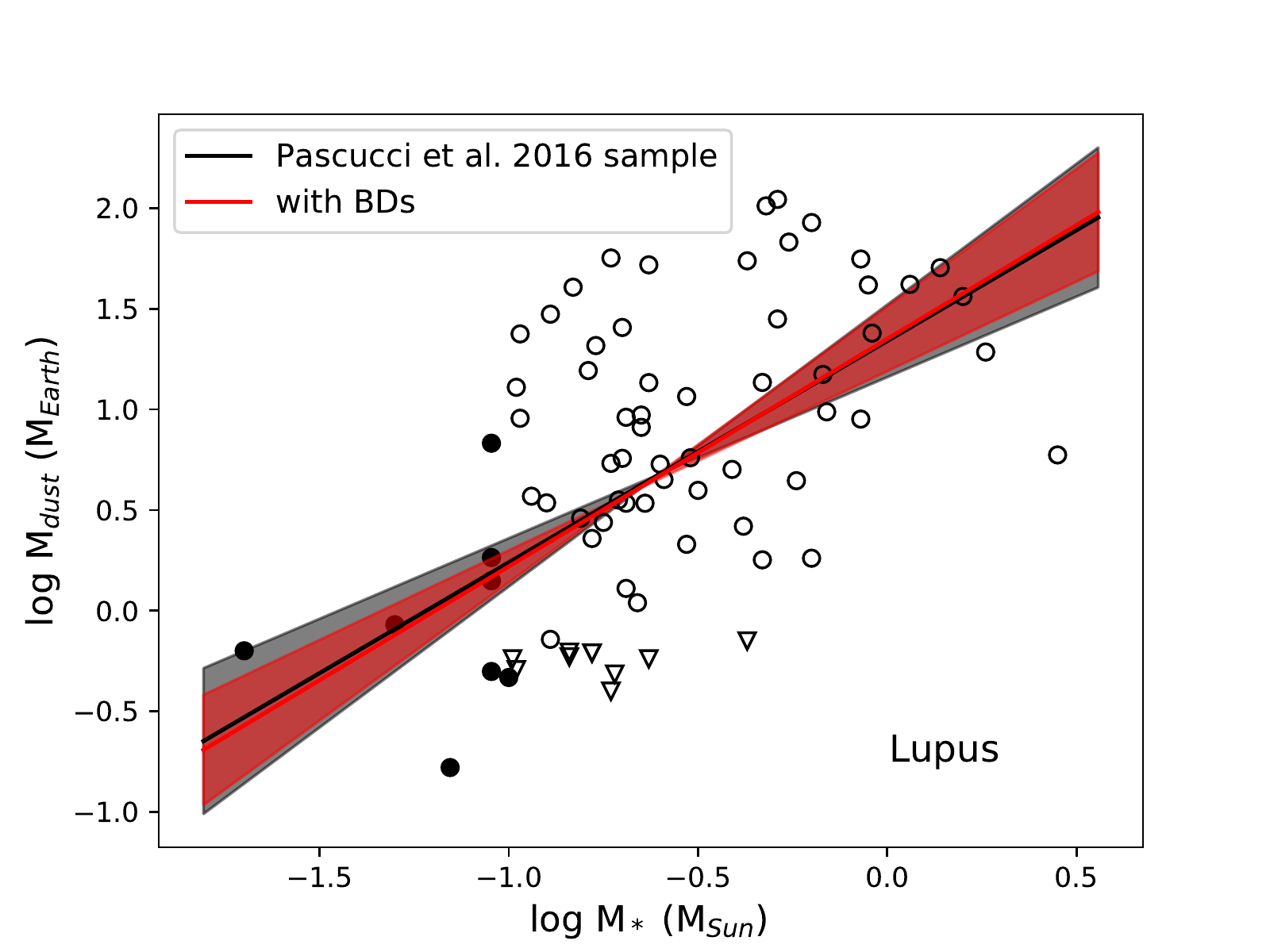}{0.34\textwidth}{(b)}
            \rightfig{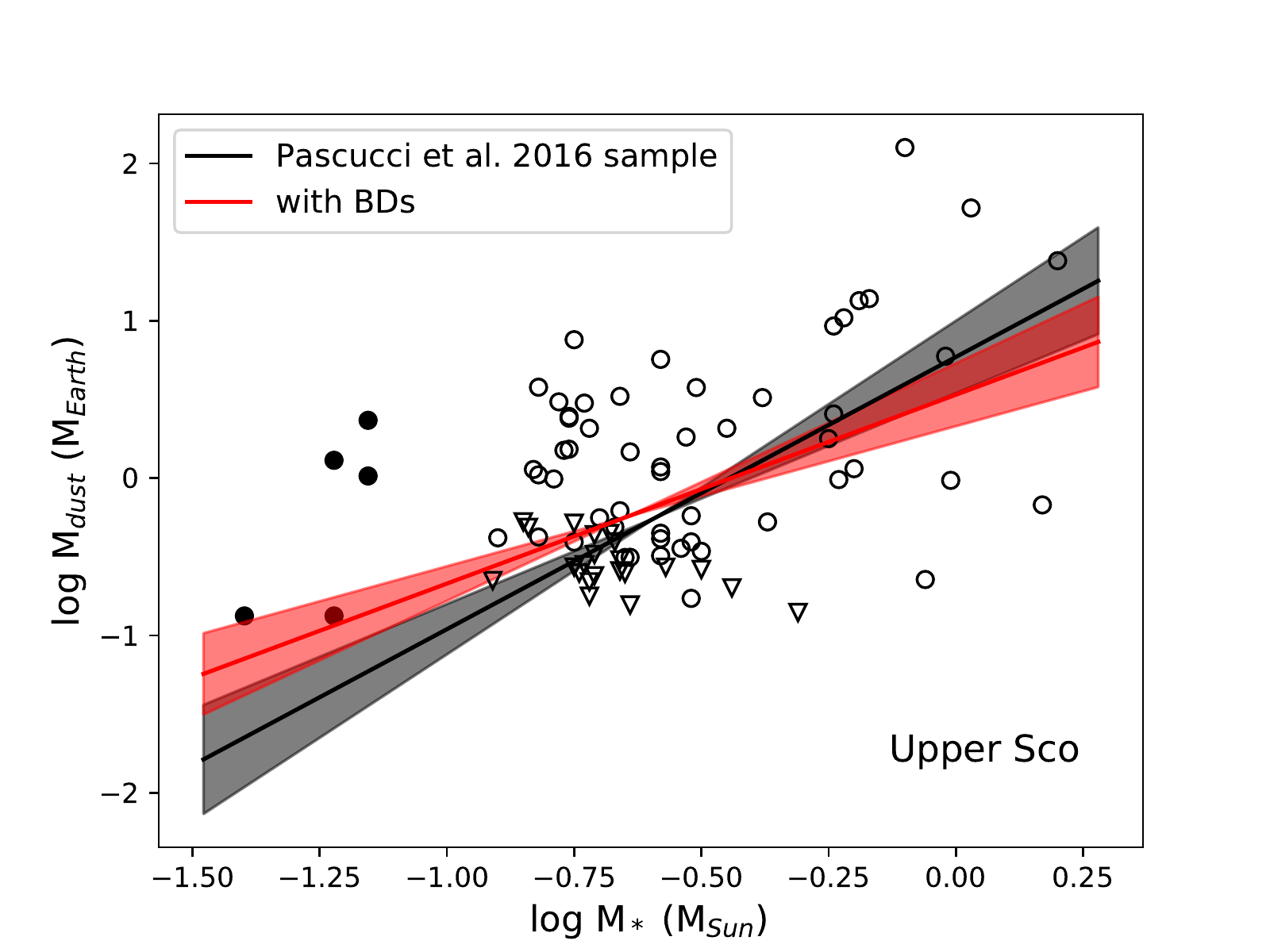}{0.34\textwidth}{(c)}}
    \caption{M$_{disk}$ versus M$_{*}$ relationships for disks in (a) Taurus, (b) Lupus, and (c) Upper Sco.  Open circles represent the sample reported in \citet{pascucci16}, filled circles represent the BDs studied in this work.  The black line and grey shaded region represent the best linear fit and uncertainties for the \citet{pascucci16} sample; the red line and red shaded region represent the best linear fit and uncertainties for the \citet{pascucci16} sample plus the BDs studied in this work.  Note that some BDs have very similar host and disk masses, so their points overlap completely.}
    \label{fig:diskmassvshostmass}
\end{figure*}

Following \citet{pascucci16}, we use the Bayesian linear fitting procedure developed by \citet{kelly07}.  This procedure accounts for the measurement uncertainties, upper limit values, and intrinsic scatter in both the disk mass and host mass values to obtain a linear relationship between disk mass and host mass in the log-log plane.  We adopt the disk mass and host mass values for objects in Taurus, Lupus, and Upper Scorpius reported by \citet{pascucci16}, who compiled data from \citet{andrews13}, \citet{ansdell16} and \citet{barenfeld16} for these three regions, respectively.  \citet{pascucci16} did not study the disk mass--host mass relationship for Ophiuchus, so we did not explore how BDs affect the relationship in that region.  For objects in both the \citet{pascucci16} sample and our sample of BDs, we adopt the disk mass obtained by our modeling.  The linear fits to this sample, with and without our sample of BDs included, are shown in Figure \ref{fig:diskmassvshostmass} and summarized in Table \ref{table:slopesandints}.

As Figure \ref{fig:diskmassvshostmass} shows, including BD disks does not substantially change the disk mass -- host mass relationship.  The two relationships (with and without our BD sample) are almost exactly the same in Lupus.  The BDs appear to slightly flatten the relationship in Taurus and Upper Sco, but the linear fits are still within the uncertainties of the relationships without the BD sample, so the effect is not significant.  Furthermore, as discussed in Section \ref{diskmasssubsection}, the disks in Upper Sco modeled here may constitute a biased sample; a more complete sample would likely include objects with lower disk masses that would bring the disk mass -- host mass relationship into closer agreement with the \citet{pascucci16} result.  A continuous M$_{disk}$ -- M$_{*}$ relationship between stellar-mass hosts and BD-mass hosts supports the theory of a single shared formation mechanism for stars and BDs \citep[e.g.,][]{hennebelle09, ricci13, ricci14, rilinger19}.

\begin{deluxetable}{c c c}
\tabletypesize{\footnotesize}
\tablecaption{M$_{dust}$ -- M$_{*}$ relationships \label{table:slopesandints}}
\tablehead{
\colhead{Region} & \colhead{$\alpha$} & \colhead{$\beta$}}
\startdata
\multicolumn{3}{c}{Without our BD sample}\\
\tableline
Taurus & 1.31 $\pm$ 0.24 & 1.13 $\pm$ 0.11\\
Lupus & 1.10 $\pm$ 0.30 & 1.34 $\pm$ 0.18\\
Upper Sco & 1.73 $\pm$ 0.39 & 0.77 $\pm$ 0.23\\
\tableline
\multicolumn{3}{c}{With our BD sample}\\
\tableline
Taurus & 0.99 $\pm$ 0.14 & 0.98 $\pm$ 0.10\\
Lupus & 1.13 $\pm$ 0.24 & 1.35 $\pm$ 0.16\\
Upper Sco & 1.10 $\pm$ 0.31 & 0.53 $\pm$ 0.20\\
\enddata
\end{deluxetable}

\section{Summary and Conclusions}\label{conclusion}
We present here the largest sample of SED models of BD protoplanetary disks to date.  We consider 49 objects located in four different star-forming regions in the Milky Way.  The DIAD radiative transfer code is used to fit models to the SEDs of these objects.  From the best-fit SED models, we obtain dust grain sizes, the degree of dust settling, and the disk mass, among other parameters, for each disk.  We compare the four star-forming regions to each other to assess how disk mass, grain growth, and dust settling vary with age.  Our results are summarized as follows:
\begin{enumerate}
    \item Though Ophiuchus is the youngest region, and thus would be expected to have the most massive disks, its disk masses are comparable to the older Taurus and Lupus regions.  This result is consistent with the TTS disk masses reported by \citet{williams19}.
    \item Upper Scorpius is the oldest region, but still contains some relatively massive disks.  This may be a selection effect, since these disks were chosen due to their high mid-IR fluxes.  However, \citet{luhmanmamajek12} note that disk lifetimes may be longer for substellar hosts, and the disk may not dissipate as quickly.
    \item All four regions show some evidence of grain growth and dust settling, indicating that dust evolution may begin to occur within 1 Myr of disk formation.
    \item Very few of the disks studied here are massive enough to form Earth-mass planetary companions according to the \citet{payne07} simulations.  Rapid planet formation and/or disk replenishment from the interstellar medium may be necessary to explain observed planetary companions to BDs.  We do not rule out the possibility that planets may form earlier and may already be present in the disks.
    \item The disk masses obtained in our modeling are consistent with previously published disk mass -- host mass relationships for stellar-mass hosts, which lends support to the theory of a shared formation mechanism between stars and BDs.
\end{enumerate}

Unbiased millimeter-wavelength studies of BD disks in these and additional star-forming regions are needed to obtain a more complete picture of how BD disks evolve over time.  The launch of the \textit{James Webb Space Telescope} will also enable detailed studies of the dust and gas in these disks, which will improve upon the disk properties presented here.

\vspace{\baselineskip}
We thank the referee for their careful reading of the manuscript and their helpful feedback. This paper utilizes the D'Alessio Irradiated Accretion Disk (DIAD) code.  We wish to recognize the work of Paola D'Alessio, who passed away in 2013. Her legacy and pioneering work live on through her substantial contributions to the field.  This work was funded by NASA ADAP 80NSSC20K0451.

\appendix
\section{SED Models}\label{appendix:models}
\renewcommand{\thefigure}{A.\arabic{figure}}
\setcounter{figure}{0}

\begin{figure*}
    \epsscale{1.2}
    \plotone{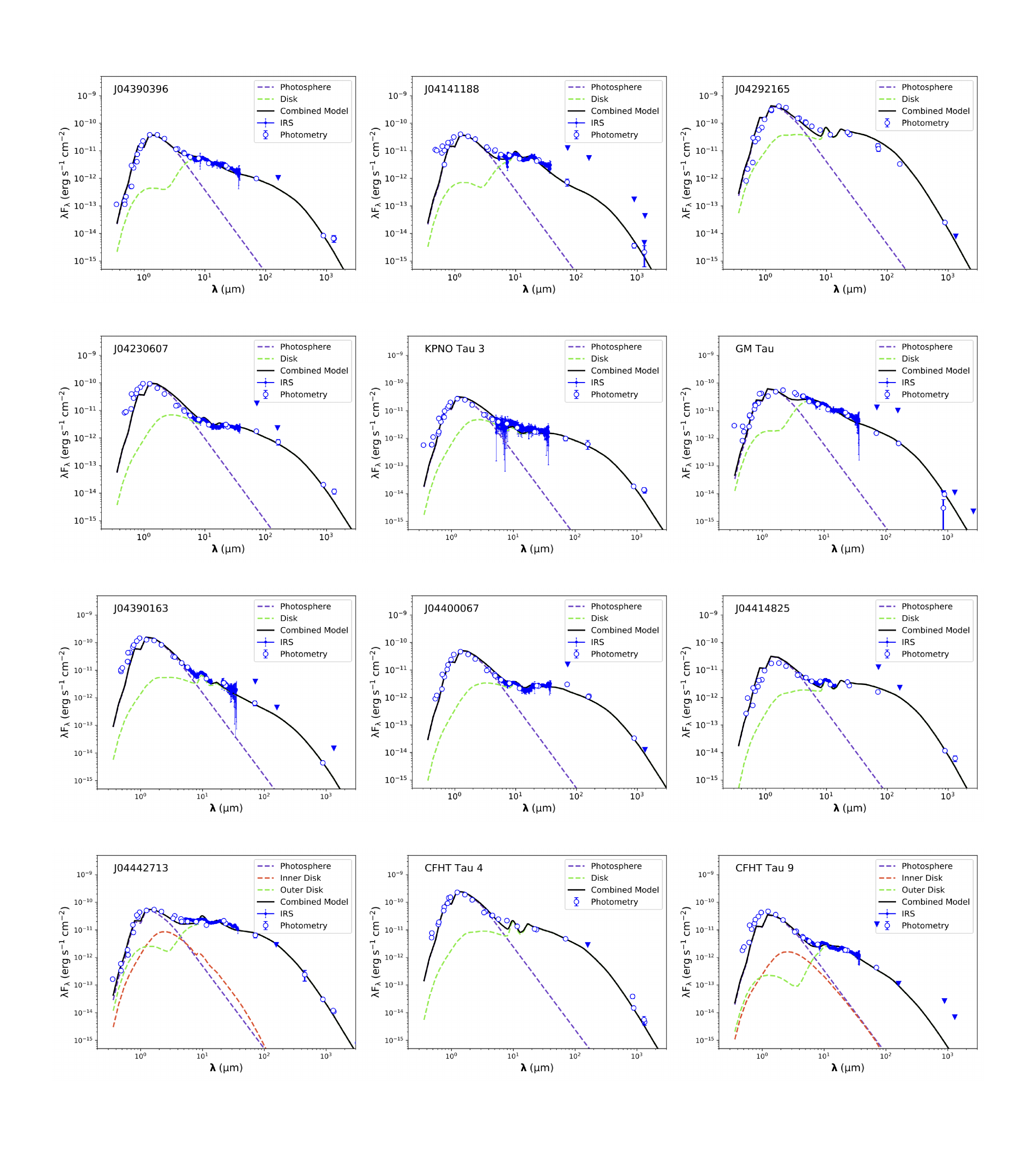}
    \caption{Spectral energy distribution models for the 23 objects in Taurus.  Filled blue triangles represent upper limits (continued on next page). (The observed photometry shown in this figure are available electronically.)}
    \label{fig:all_tau_seds}
\end{figure*}

\begin{figure*}
    \epsscale{1.2}
    \figurenum{A.1}
    \plotone{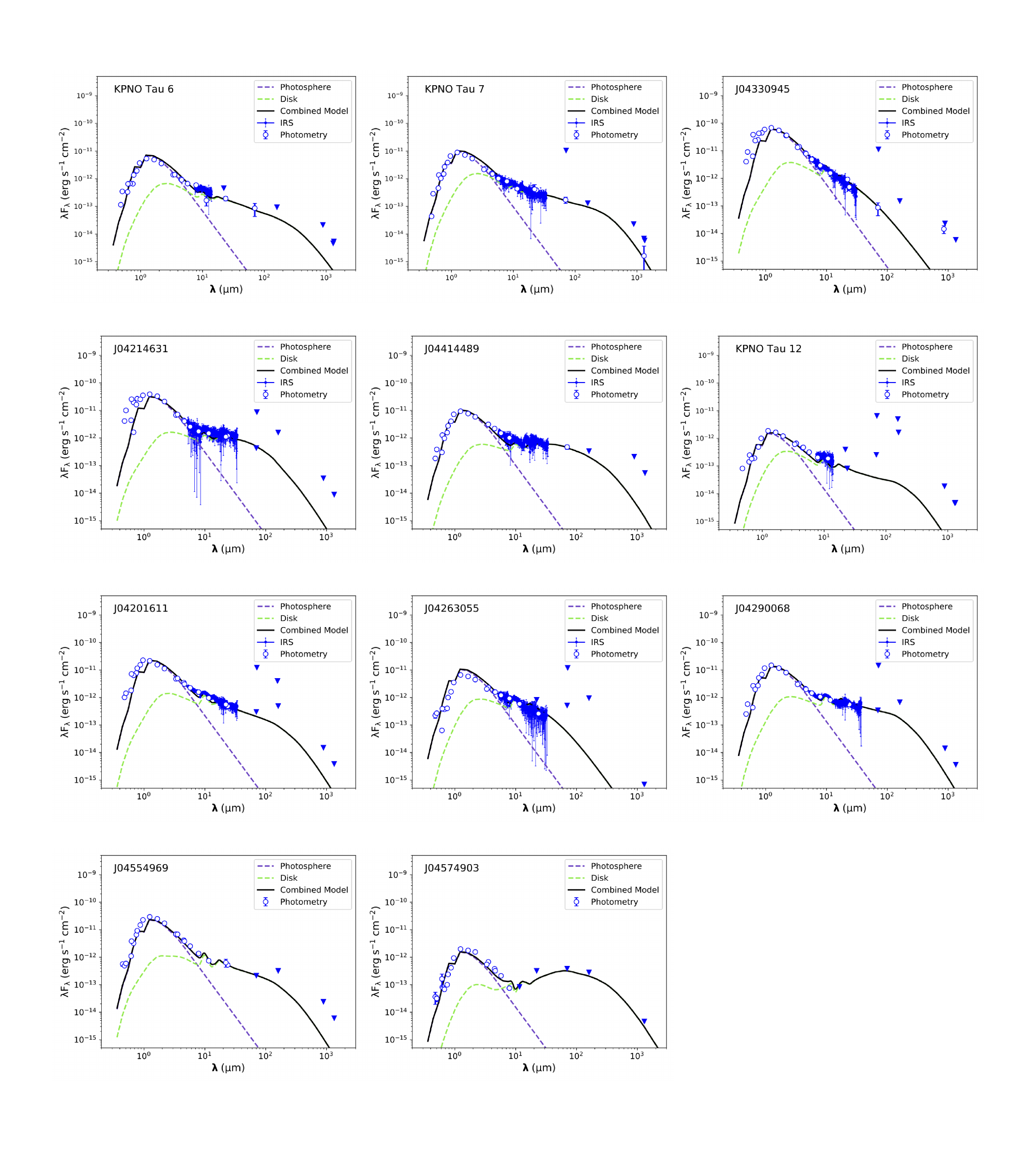}
    \caption{Continued: Spectral energy distribution models for the 23 objects in Taurus. Filled blue triangles represent upper limits. (The observed photometry shown in this figure are available electronically.)}
\end{figure*}

\begin{figure*}
    \epsscale{1.2}
    \plotone{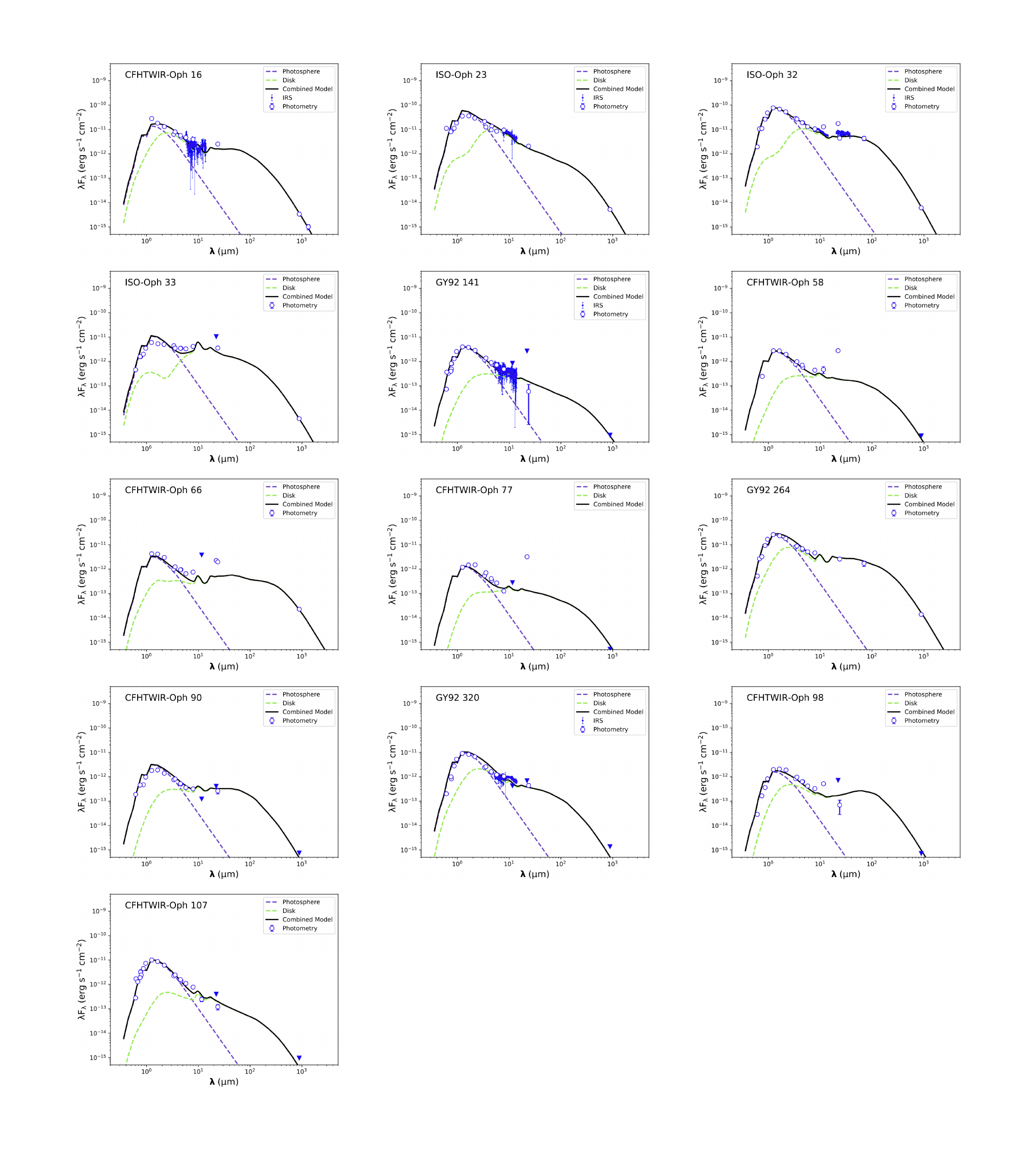}
    \caption{Spectral energy distribution models for the 13 objects in Ophiuchus. Filled blue triangles represent upper limits. (The observed photometry shown in this figure are available electronically.)}
    \label{fig:all_oph_seds}
\end{figure*}

\begin{figure*}
    \epsscale{1.2}
    \plotone{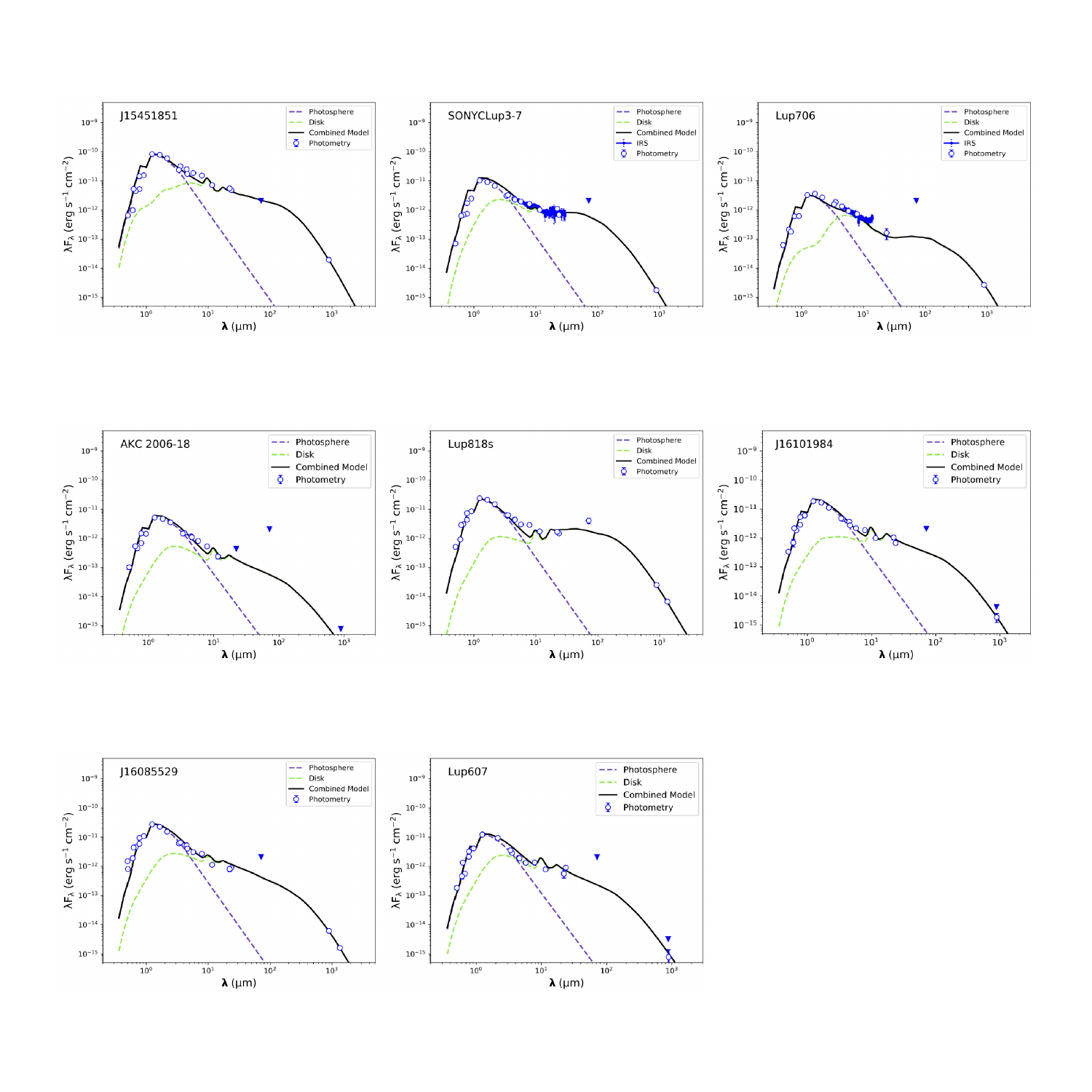}
    \caption{Spectral energy distribution models for the 8 objects in Lupus. Filled blue triangles represent upper limits. (The observed photometry shown in this figure are available electronically.)}
    \label{fig:all_lup_seds}
\end{figure*}

\begin{figure*}
    \epsscale{1.2}
    \plotone{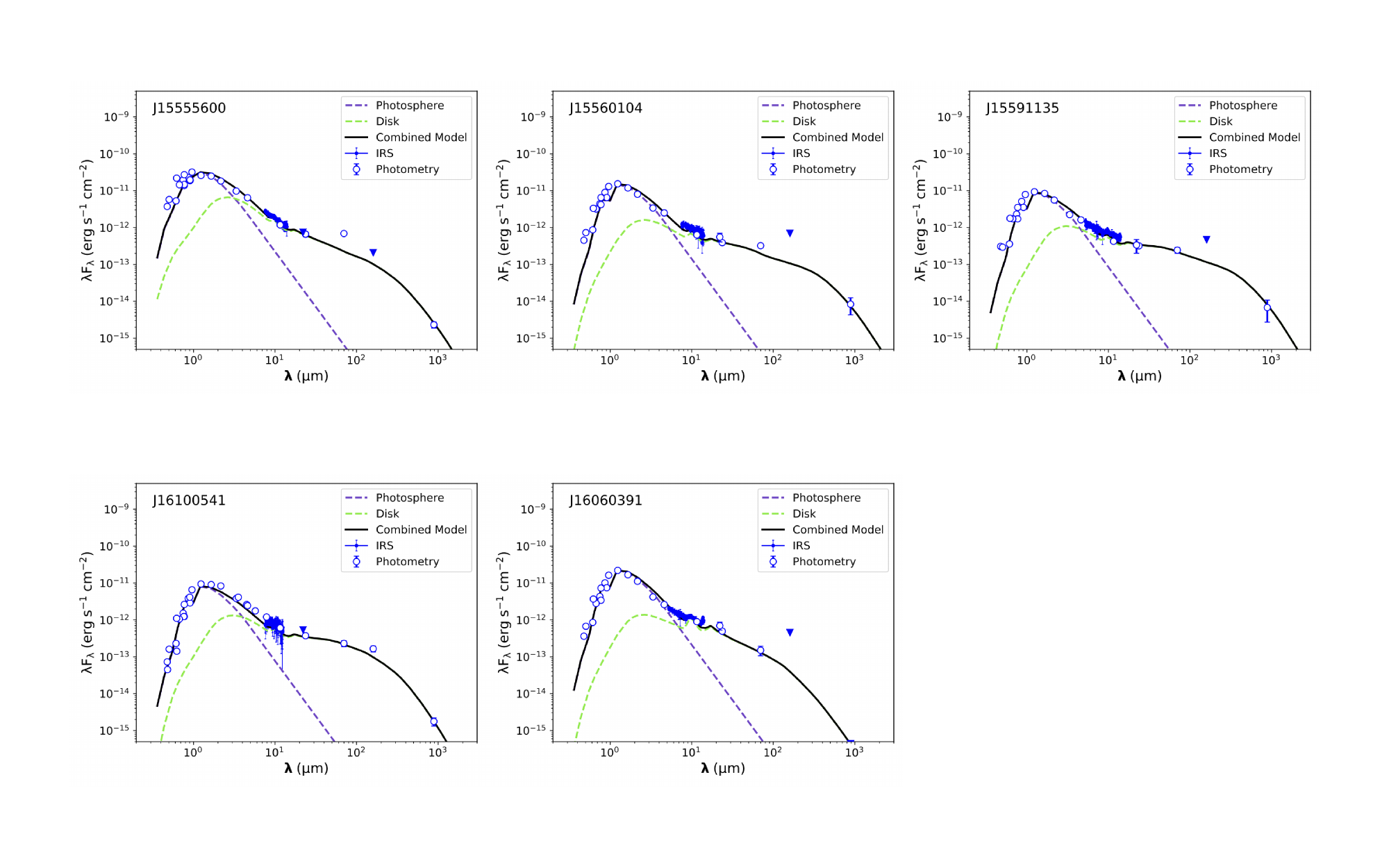}
    \caption{Spectral energy distribution models for the 5 objects in Upper Scorpius. Filled blue triangles represent upper limits. (The observed photometry shown in this figure are available electronically.)}
    \label{fig:all_sco_seds}
\end{figure*}

\bibliographystyle{apj}
\bibliography{refs}

\end{document}